\documentclass[twocolumn,showpacs,showkeys,preprintnumbers,superscriptaddress,amsmath,floatfix,amssymb,secnumarabic,nofootinbib]{revtex4-1}
\usepackage[colorlinks=true]{hyperref}
\usepackage{graphicx}
\usepackage{dblfloatfix}    
\usepackage{placeins}
\usepackage{braket}
\usepackage{amsmath}
\usepackage{physics}
\usepackage{epsfig}
\usepackage{float}
\usepackage{nicefrac}
\usepackage{xcolor}
\usepackage{color}
\usepackage{bm}
\usepackage{subfigure}
\usepackage{chngcntr}
\usepackage[normalem]{ulem}
\usepackage{multirow}

\def\({\left(}
\def\){\right)}
\def\[{\left[}
\def\]{\right]}

\newcommand{\K}{\, {\rm K}}

\newcommand{\beq} {\begin{eqnarray}}
\newcommand{\eeq} {\end{eqnarray}}

\begin{document}
\sloppy

\title{Plasmon excitations in half-filled graphene: A Comparative study between Quantum Monte Carlo and Random Phase Approximation }

\author{Adrien~Reingruber}
\email{adrien.reingruber@uni-wuerzburg.de}
\affiliation{Institut f\"ur Theoretische Physik und Astrophysik, Universit\"at W\"urzburg, 97074 W\"urzburg, Germany}

\author{Maksim~Ulybyshev}
\email{maksim.ulybyshev@uni-wuerzburg.de}
\affiliation{Institut f\"ur Theoretische Physik und Astrophysik, Universit\"at W\"urzburg, 97074 W\"urzburg, Germany}

\author{Kitinan~Pongsangangan}
\email{kitinan.pon@mahidol.ac.th}
\affiliation{Department of Physics, Faculty of Science, Mahidol University, Bangkok 10400, Thailand}

\begin{abstract}

Transport properties of strongly correlated materials have contributions from quasiparticle excitations such as electrons and holes as well as emerging collective excitations such as plasmonic sound-like modes which are sustained by interactions. It was previously shown in Ref.\cite{Kitinan2022b} that the thermal excitation of the long-lived plasmons in graphene provides a substantial contribution to heat and momentum transport in the interaction-dominated regime. 
Detailed information on these excitations is therefore necessary for the understanding of hydrodynamic transport with quantitative precision. On the other hand, dynamics of graphene plasmons is usually studied using perturbation theory within the Dirac-cone approximation, thus neglecting the effects of a finite Brillouin zone and higher-order perturbative corrections. Both these effects can be however significant for strong-interacting systems including free-standing graphene where the effective coupling constant can reach values up to $\alpha = \frac{e^2}{\hbar v_{\mathrm{F}}} = 2$. Consequently, in this paper, we studied the behavior of plasmons in half-filled free standing graphene using unbiased Quantum Monte Carlo (QMC) calculations. We confirm the existence of well-defined resonance peaks for plasmons around the $\Gamma$-point. Comparison with the Random-phase-approximation (RPA) calculation for the honeycomb lattice shows a qualitative agreement with QMC.
However, RPA yields more stable plasmon modes, while QMC shows enhanced broadening due to non‑perturbative interaction effects naturally included in the simulations.
To account for this effect, we generalize the calculation of Lattice RPA by dressing the fermionic propagator by a constant lifetime. As a result, the plasmon frequencies are shifted towards higher energies and the spectrum is broadened, yielding a better agreement with QMC. Our findings highlight the need to account for both a finite Brillouin zone and strong interaction effects when developing theories of electronic transport in free-standing graphene.

\end{abstract}

\maketitle

\section{\label{sec:Intro}Introduction}

Collective excitations in condensed matter systems have been a subject of theoretical and experimental investigations for decades. These collective modes constitute one of the fundamental excitations in interacting systems and are absent in the non-interacting counterparts. In a system of electrons with the long-range Coulomb interaction, the most well-known collective excitation is the plasmon \cite{Pines1999,Bohm1953,Vlasov_1968,Lindhard1953,Stern1967}, which manifests itself as a resonance peak in the density-density correlation function \cite{Giuliani2005} and is usually experimentally observable using electron energy loss spectroscopy \cite{Liu2008,Liou2015,Shin2011}. 

In addition to providing a restoring force that supports the emergence of collective charge density oscillations or plasmons, the Coulomb interactions also mediate momentum-conserving electron–electron collisions, which can push the system into the hydrodynamic regime. The hydrodynamic transport exhibits viscous behavior described by the Navier-Stokes-type equation \cite{Gurzhi68,deJong95,Damle1997,Muller2009,Lucas_2018,Ho18,Narozhny2019}. First described by Gurzhi in the 1960s\cite{Gurzhi68}, when momentum conserving electron–electron scattering dominates over diffusive scattering (defects, phonons), it enables a collective movement of charge carriers with non-trivial current-flow profiles \cite{Levitov2016}. Evidence for hydrodynamic electron flow in materials dates back to the early experiments of Molenkamp et al. in 1995\cite{deJong95}, yet only recent progress in sample preparation and imaging has enabled the direct visualization of hydrodynamic flow patterns in high-quality graphene
\cite{Bandurin2018, Sulpizio2019,Ku2020,Jenkins2022}.
The Dirac quasiparticles in graphene\cite{Katsnelson_2020} form an ideal platform for exploring hydrodynamic transport as the density-of-states is vanishingly small close to the charge-neutrality point and, consequently, the Coulomb interaction remains unscreened. The electronic fluid in charge-neutral graphene was suggested to approach the minimal possible value of the shear-viscosity-to-entropy-density ratio\cite{Muller2009}, a limit characteristic of a broad class of strongly correlated QFTs \cite{Kovtun2005,Karsch2008}.

Recent Quantum Monte Carlo (QMC) studies based on a microscopic quantum lattice-model of charge-neutral graphene with realistic long-range Coulomb interactions have shown the emergence of the hydrodynamic current flow profile \cite{QMCHydro2025}. This raises a question: Can QMC capture another manifestation of the Coulomb interaction, i.e., collective excitations like plasmons? Such check would be especially interesting at half filling, where the existence of plasmons as well-defined quasiparticle peaks in electron density channel is \textit{a priori} not obvious due to the zero average electronic density. Although their existence was theoretically predicted a decade ago in Ref.\cite{Sarma2013}, the very first glimpse of plasmons in pristine graphene has only been available recently in a terahertz-spacetime-metrology experiment \cite{xu2026plasmondynamicsgraphene}. In addition, the kinetic-equation approach suggested that plasmons have a sizable contribution to thermal and momentum transport \cite{Kitinan2022a,Kitinan2022b}. Information on this collective excitation  is therefore necessary for a future comparison of the QMC simulation and the Boltzmann-equation-based calculation of thermal conductivity and shear viscosity of graphene in the interaction-dominated regime. Furthermore, plasmon relaxation time is closely related to electron and hole scattering rate via the conservation laws \cite{BaymKadanoff1961}. Thus, understanding the plasmon lifetime will also serve as a hint for the fermions self-energies and, consequently, electron-electron collision time.

 Plasmons in graphene are well described in the Dirac theory within the Random phase approximation (RPA) according to which plasmons do not interact with each other and only decay into a particle-hole pair due to the Landau damping process \cite{Pines1952,Bohm1953,Lindhard1953,Stern1967,Wunsch_2006,Liu2008,Sarma2013}. However, the influence of the underlying discrete honeycomb lattice \cite{Stauber2010,Hill2009} and strong electron–electron correlations remains largely unexplored, especially at the charge neutrality point. In fact, plasmon's spectrum and lifetime broadening are sensitive to many-body dynamics \cite{Macdonald2011,Levitov2013}. In this case, exact lattice-model simulations provide a powerful route forward: they can deliver clean, non-perturbative predictions for both plasmon excitations and transport coefficients directly from the microscopic Hamiltonian, while simultaneously establishing a uniquely controlled environment in which approximate analytic approaches—kinetic equations or RPA—can be tested quantitatively against unbiased quantum Monte Carlo results\cite{Ulybyshev2023,QMCHydro2025}. Lattice QMC \cite{Blankenbecler81,White89, Sorella_1992,Assaad08_rev} simulations are able to not only provide exact reference data but, in regimes that are hard to probe experimentally, effectively complement or even substitute measurements when evaluating the quantitative reliability of kinetic-theory and RPA predictions.

In the present study, we investigate the emergence of plasmon excitations in interacting graphene directly on the honeycomb lattice. Using large-scale, sign-problem-free QMC simulations at charge neutrality\cite{Sorella_1992,Assaad08_rev, Ulybyshev13, Hohenadler14, Jiménez2021} , we retain the full long-range Coulomb interaction and treat electron correlations exactly. The emergence of clear gaps in other ordered-state (spin- and charge-density-wave) excitations, contrasted with a persistent sharp gapless mode in the density–density channel, unambiguously identifies this mode as the plasmon and simultaneously confirms the resolution and reliability of our QMC simulations. 
The results are compared with lattice‑RPA (LRPA) calculations on the honeycomb lattice with the long-range Coulomb potential, providing the estimates for the effects, not captured by RPA with the Dirac-cone approximation. 

 \begin{figure}[]
  \centering
\includegraphics[width=0.45\textwidth , angle=0]{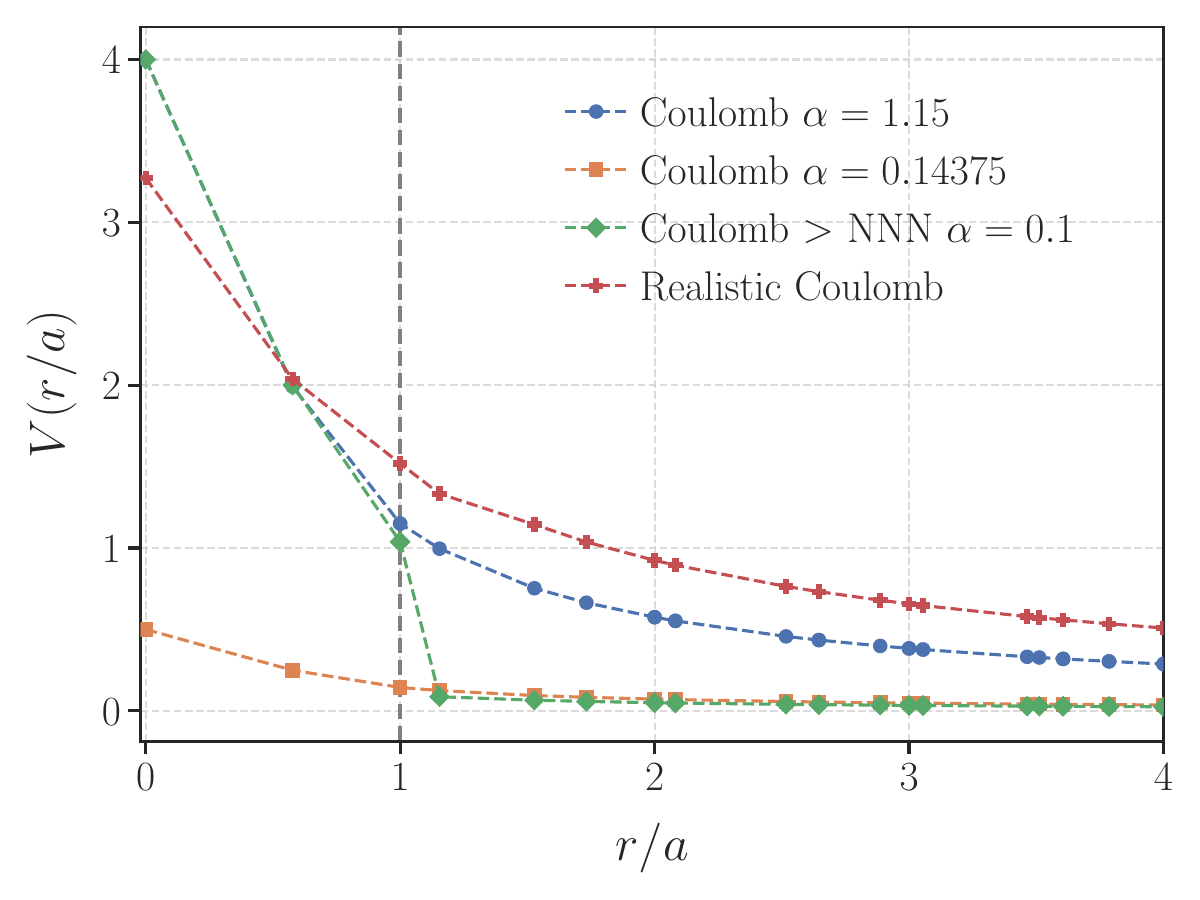}
        \caption{Coulomb potential of different setups used in our QMC simulations}
  \label{fig:Coulomb}
\end{figure}

\section{\label{sec:model}Model}

We consider the Hamiltonian of the interacting electrons in graphene given by
\begin{equation}
    \hat H[\hat{a}_{x,\sigma},\hat{a}^\dag_{x,\sigma}]=  H_{\text{tb}}[\hat{a}_{x,\sigma},\hat{a}^\dag_{x,\sigma}] + H_{\text{int}}[\hat{a}_{x,\sigma},\hat{a}^\dag_{x,\sigma}],
    \label{eq:basic_H}
\end{equation}
where 
\begin{equation}
H_{\text{tb}}[\hat{a}_{x,\sigma},\hat{a}^\dag_{x,\sigma}]= -t \sum_{\sigma}\sum_{<x,y>} \hat a^\dag_{x, \sigma} \hat a_{y, \sigma} + h.c, 
\label{eq:Htb}
\end{equation}
is the tight-binding model on the honeycomb lattice with the hopping parameter $t$ and
\begin{equation}
H_{\text{int}}[\hat{a}_{x,\sigma},\hat{a}^\dag_{x,\sigma}] = \frac{1}{2} \sum_{x, y} \hat q_x V_{x,y} \hat q_y
\end{equation}
accounts for the Coulomb interaction.
Here $a^\dag_{x, \sigma}$ is the electron creation operator for spin $\sigma$ with the index $x=(\vec r,\eta)$ including the 2D spatial coordinates of the unit cell $\vec r$ and the sublattice index $\eta=0,1$. Nearest neighbor sites are denoted by  $<x,y>$. The Coulomb interaction describes the interaction between the charge densities which are given by $\hat q_x = \sum_{\sigma=\uparrow, \downarrow} \hat a^\dag_{x, \sigma} \hat a_{x, \sigma} - 1$. The electron-electron interaction consists of the Hubbard on-site interaction of strength $U$ and the Coulomb tail starting from the interaction with nearest-neighbors.  From now on, we express all dimensional quantities (e.g. temperature) in the units of hopping amplitude $t$. 

In our QMC computations we consider three interaction setups, visualized in Fig.~\ref{fig:Coulomb}.
\begin{enumerate}
    \item Full Coulomb setup:
   
The Coulomb interaction follows equation 
\begin{equation}
V_{x,y} = \frac{\alpha t}{|\vec{x}-\vec{y}|}
\label{eq:Coulomb_law}
\end{equation}
starting from the nearest neighbors. $V_{x,y}$ is interaction potentials (expressed in units of hopping), the distance $|\vec{x}-\vec{y}|$ is in the units of the lattice unit cell length, and $\alpha$ is a dimensionless constant.  In the strong-interaction case we set $\alpha = 1.15$, which yields a nearest-neighbor interaction $V = 2.0\,t$ and an on-site interaction $U = 4.0\,t$. The weak-interaction case is obtained by uniformly rescaling both the on-site interaction and all nonzero Coulomb interactions by a factor of $1/8$.
    
    \item Coulomb $>$ NNN setup:
    The on‑site, nearest‑neighbor, and next‑to‑nearest‑neighbor interactions are held fixed, whereas all longer‑range terms follow the Coulomb law \eqref{eq:Coulomb_law} and are uniformly rescaled to $\alpha = 0.1$. We refer to this scheme as " Coulomb $>$ NNN" in Fig.~\ref{fig:Coulomb}. Although somewhat artificial, it highlights qualitative differences between interaction types, since it allows us to clearly distinguish the regimes with dominant long-range Coulomb tail or inversely, with dominant short-range interactions. This setup also loosely models graphene on a substrate, where screening primarily suppresses the long‑range Coulomb tail.
    \item Realistic Coulomb setup:
    We also include some simulations made with realistic Coulomb interactions corresponding to suspended graphene. These interactions are taken from \cite{Wehling2011} where screening by electrons on $\sigma$-orbitals is taken into account at intermediate distances, and the long range asymptotics of the Coulomb tail corresponds to the vacuum dielectric permittivity. Recently, these parameters were validated by comparing large-scale QMC simulations to experimental data for the renormalized Fermi velocity $v_F$ in  \cite{Ulybyshev2023}.
\end{enumerate}

\section{\label{sec:plasmons}Plasmons}

The plasmon resonance is obtained from the Euclidean time density-density correlator given by
\begin{eqnarray}
C_q(\vec k, \tau)=\frac{1}{N_{cells}}\sum_{\vec r_1,\vec r_2} e^{-i \vec k (\vec r_1-\vec r_2)}\langle \hat q_{\vec r_1}(0) \hat q_{\vec r_2}^\dag(\tau)\rangle,
\label{eq:corr}
\end{eqnarray}
where $N_{cells}$ is the number of unit cells in our finite lattice, and the charge density operator at the lattice site $\vec{r}$ is a sum of density at both sublattices, i.e., $\hat{q}_{\vec{r}} = \hat{q}_{(\vec{r},0)} +\hat{q}_{(\vec{r},1)}$. The Euclidean time correlator is subsequently converted to real-frequency spectral functions using analytical continuation \cite{Sandvik98,Beach04a,SHAO20231}. 

We start from the usual Trotter decomposition:
\begin{eqnarray}
Z=\Tr{\prod_{t} e^{-\Delta t H_{\text{tb}} } e^{-\Delta t H_{\text{int}} }  },
\label{eq:trotter}
\end{eqnarray}
with $\Delta t$ being the step in Euclidean time, such that $\Delta t N_t=\beta = 1/T$, where the temperature $T$ takes two values: $0.25t$ and $0.1t$. 
Subsequently, we introduced a real scalar field $\phi_{x}(t)$ to decouple the electron-electron interaction using a Hubbard-Stratonovich transformation. This gives the partition function for the Hamiltonian Eq.\eqref{eq:basic_H} reading as
\begin{eqnarray}
Z &=& \int \prod_{t, x} d \phi_x(t) d \bar{\psi}_x(t) d \psi_x(t) ~  e^{-\Delta t \sum_t \hat H_{tb}[\psi_{x,\sigma}(t),\bar{\psi}_{x,\sigma}(t)]}  \nonumber \\ &&  e^{i \Delta t\sum_{x,t} \hat q_x (t)\phi_x(t) -\frac{\Delta t}{2} \sum_{t, x,y} \phi_x(t) \left ( {D^{(0)}}^{-1}\right)_{x,y} \phi_y(t) } .
\end{eqnarray}
Here $\psi_{x,\sigma}(t)$ and $\bar{\psi}_{x,\sigma}(t)$ are Grassmann fields associated with the electron annihilation and creation operators, respectively. We have also introduced the bare bosonic propagator ${D_{x,y}^{(0)}}^{-1}=V_{x,y}^{-1}$.  Grassmann fields can be integrated out using a Gaussian integral. This gives an effective theory of bosons which can be solved exactly using a QMC simulation:
\begin{eqnarray}
Z &=& \int \prod_{t, x} d \phi_x(t) e^{-\frac{\Delta t}{2} \sum_{t,x,y} \phi_x(t) \left ( {D^{(0)}}^{-1}\right)_{x,y} \phi_y(t)} \nonumber \\ && \hspace{1.5cm} e^{  \Tr \ln \left( \Delta t G_{0xy\sigma}^{-1}(t) - i\Delta t \phi_{x}(t) \right)   }
\label{eq:bosontheory}
\end{eqnarray}
Here we define $G^{-1}_0$ from the tight-binding model as $H_{\text{tb}}[\psi_{x,\sigma}(t),\bar{\psi}_{x,\sigma}(t)] = \sum_{t,x,y,\sigma}\bar{\psi}_{x,\sigma}(t) G^{-1}_{0xy\sigma}(t) \psi_{y,\sigma}(t)$.

The density-density correlator Eq.\eqref{eq:corr} can be expressed in terms of the exact boson propagator,
\begin{eqnarray}
     D_{z1_, z_2}(\tau)=\langle \phi_{\vec z_1}(0) \phi_{z_2}(\tau)\rangle,
     \label{eq:boson_prop}
\end{eqnarray}
by using the derivatives of partition function with respect to $\phi_x(0)$ and $\phi_y(\tau)$ fields to extract charge operators from the exponents under the trace. These derivatives can then be transferred to derivatives of the bosonic exponent with respect the same field using integration by parts: 
\begin{eqnarray}
    \langle \hat q_{\vec x}(0) \hat q_{y}(\tau)\rangle  = \\
    - \sum_{z_1, z_2} \left ( {D^{(0)}}^{-1}\right)_{x,z_1}  D_{z1_, z_2}(\tau) \left ( {D^{(0)}}^{-1}\right)_{z_2,y},  \nonumber 
\label{eq:connection}
\end{eqnarray}
which leads us to the following expression in momentum space
\begin{eqnarray}
    C_q(\vec k, \tau) = - {D^{(0)}}^{-1}(\vec k)  D(\vec k, \tau)  {D^{(0)}}^{-1}(\vec k).
\label{eq:connectionK}
\end{eqnarray}
Therefore, plasmon's energy dispersion and lifetime can be extracted from either the poles of the density-density correlator or the bosonic propagator.

Transition to real frequencies is done using the Green-Kubo relations
\begin{eqnarray}
 C_q(\vec k, \tau)=\int_0^{\infty}  \frac{ \Im C_q(\vec k, \omega) }{\pi}  \frac{\cosh(\omega (\tau-\beta/2))}{\sinh(\omega \beta/2)}     d\omega,
 \label{eq:GK}
\end{eqnarray}
which are solved using stochastic analytical continuation technique \cite{Beach04a,Sandvik98,SHAO20231}. Subsequently, we use equations \eqref{eq:boson_prop} -  \eqref{eq:GK} to make a connection with the Lattice RPA computation for plasmons. In particularly, the RPA spectral function is connected to the QMC one via 
\begin{eqnarray}
\Im C_q(\vec k, \omega)  =  - {D^{(0)}}^{-1}(\vec k)  A(\omega,k)  {D^{(0)}}^{-1}(\vec k).
 \label{eq:QMC_RPA_connection}
\end{eqnarray}

\section{Lattice RPA}
While QMC can solve the boson propagator exactly, the Boltzmann factor exponent in the partition function Eq. \eqref{eq:bosontheory} is kept up to quadratic order in $\phi$ within RPA. In order to make a closer comparison with QMC, which is made on microscopic hexagonal lattice, we formulate lattice-RPA, which uses exactly the same microscopic Hamiltonian \ref{eq:Htb}.  The boson propagator in RPA is given by
\begin{equation}
D(\vec{k},\omega) = \frac{1}{{{D^{(0)}}^{-1}}(\vec{k})-\Pi^{(0)}(\vec{k},\omega)},
\label{eq:D_RPA}
\end{equation}
where $\Pi^{0}(\vec{k},\omega)$ is the polarization bubble, which is computed with the full graphene energy spectrum using the Lindhard formula 
\begin{eqnarray}
&&\Pi^{(0)}(\vec{k},\omega)  = \nonumber \\ && \hspace{0.6cm} N \sum_{\lambda,\lambda'=\pm1}\int \frac{d\vec{q}}{(2\pi)^2}  \frac{\mathcal{F}_{\lambda\lambda'}(\vec{k},\vec{q})\left(f_\lambda(\vec{q})-f_{\lambda'}(\vec{k}+\vec{q})\right)}{\omega+i0^{+}+\epsilon_\lambda(\vec{q})-\epsilon_{\lambda'}(\vec{k}+\vec{q})}. \nonumber \\
\label{eq:Lindhard}
\end{eqnarray}
Here $f_\lambda(\vec{k})$  is the fermi-Dirac function evaluated at energy $\epsilon_{\lambda}(\vec{k})$. It is given by $\epsilon_{\lambda}(\vec{k})= \lambda t |\phi_{\vec{k}}|$ with $\lambda$ being the band index and $\phi_{\vec{k}} = \sum_{i=1}^{3} e^{-i(\vec{\delta}_i-\vec{\delta}_3)\cdot\vec{k}}$. We choose $\vec{\delta}_i$, the three nearest neighbor vectors, to be $\vec{\delta}_1=\frac{a_{CC}}{2}(-1,\sqrt{3})$, $\vec{\delta}_2=\frac{a_{CC}}{2}(-1,-\sqrt{3})$ and $\vec{\delta}_3=a_{CC}(1,0)$ with $a_{CC}$ being the nearest carbon-carbon distance. The wavefunction overlap $\mathcal{F}_{\lambda\lambda'}(\vec{k},\vec{q}) = \frac{1}{2} \left( 1 + \lambda\lambda' \Re\left[{e^{-i\vec{k}\cdot\vec{\delta}_3}}\frac{\phi_{\vec{q}} \phi^*_{\vec{k}+\vec{q}}}{|\phi_{\vec{q}}|| \phi^*_{\vec{k}+\vec{q}}|}\right]\right)$ and, as we integrate over the first Brillioun zone, here $N=2$ accounts only for spin degeneracy. The integral in \ref{eq:Lindhard} is evaluated numerically. For simplicity, we only included pure Coulomb tail \ref{eq:Coulomb_law} in the free bosonic propagator in \ref{eq:D_RPA}. We should stress again that this calculation is beyond the Dirac-cone approximation usually assumed in the literature.

\begin{figure}[h]
  \centering
\includegraphics[width=0.4\textwidth , angle=0]{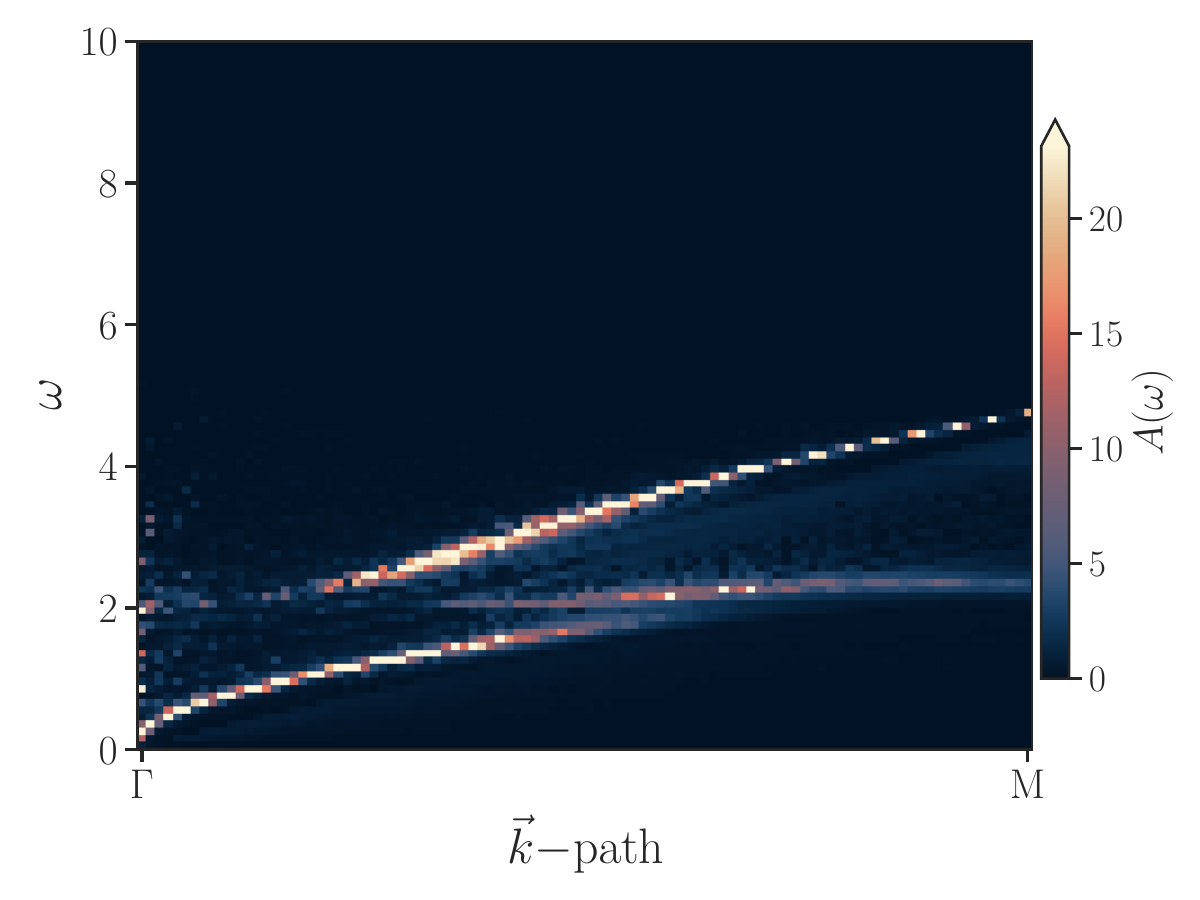}
         \caption{Spectral function, $ A(\omega,k)$, numerically evaluated from LRPA for graphene with interaction strength $\alpha=2$ and the hopping parameter to temperature ratio $t/T=1.$  }
    \label{fig:LRPA}
\end{figure}

As an example, Fig.~\ref{fig:LRPA} shows the RPA plasmonic spectral function for graphene with interaction strength $\alpha=2$. In the long-wavelength limit, we find clear plasmon peaks which follow a square-root dispersion relation as to be expected for any two dimensional systems. In addition to that, at higher energy (around $\omega/T = 2-5$), we find another plasmon branch resulting from interband transitions. It should be noted that this high-energy plasmon mode was previously discussed in Ref.\cite{Levitovplasmon2019} for twisted bilayer graphene.

  \begin{figure}[]
   \centering
   \subfigure     { \label{fig:T0.1_large_alpha_1el}\includegraphics[width=0.35\textwidth]{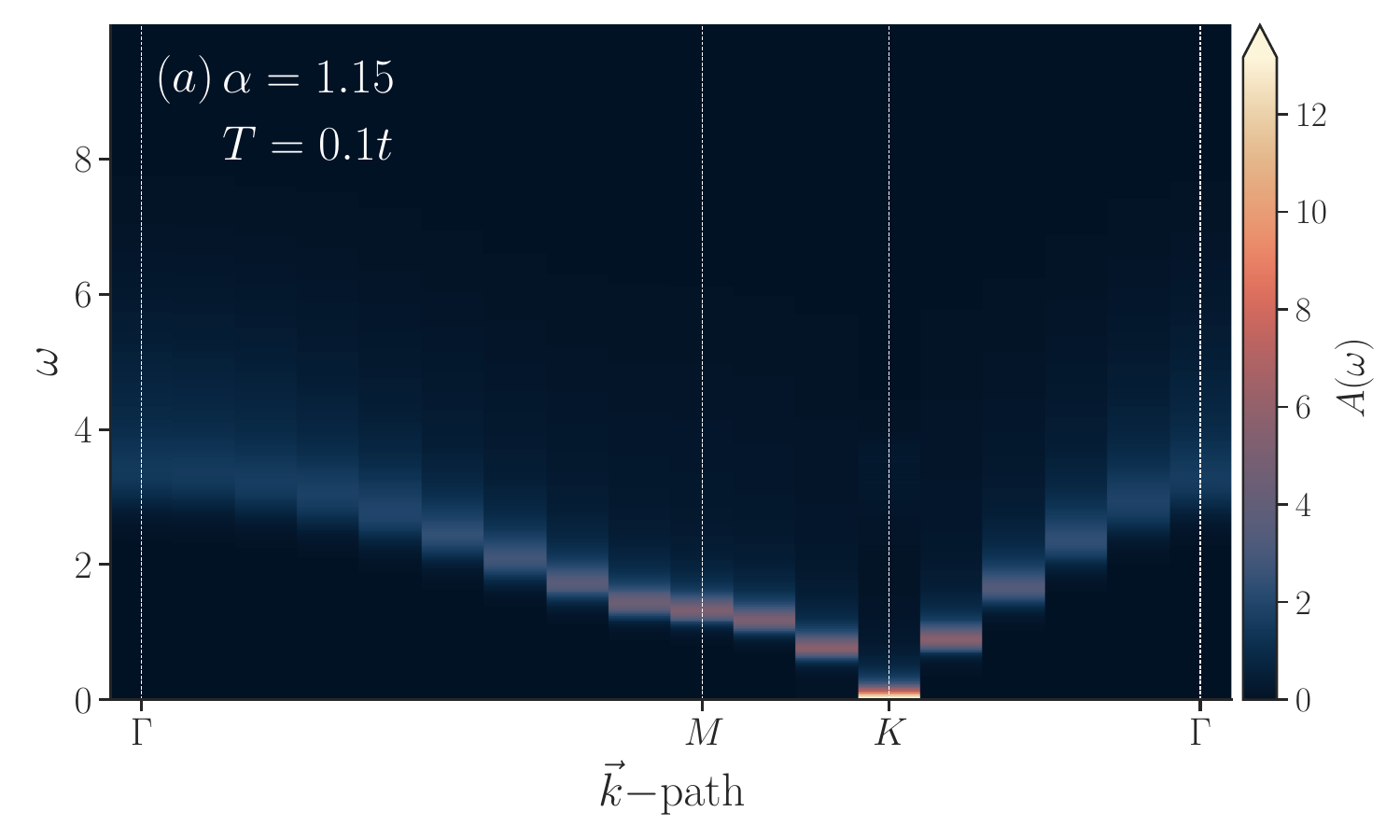}}
    \subfigure     { \label{fig:T0.1_small_alpha_1el}\includegraphics[width=0.35\textwidth]{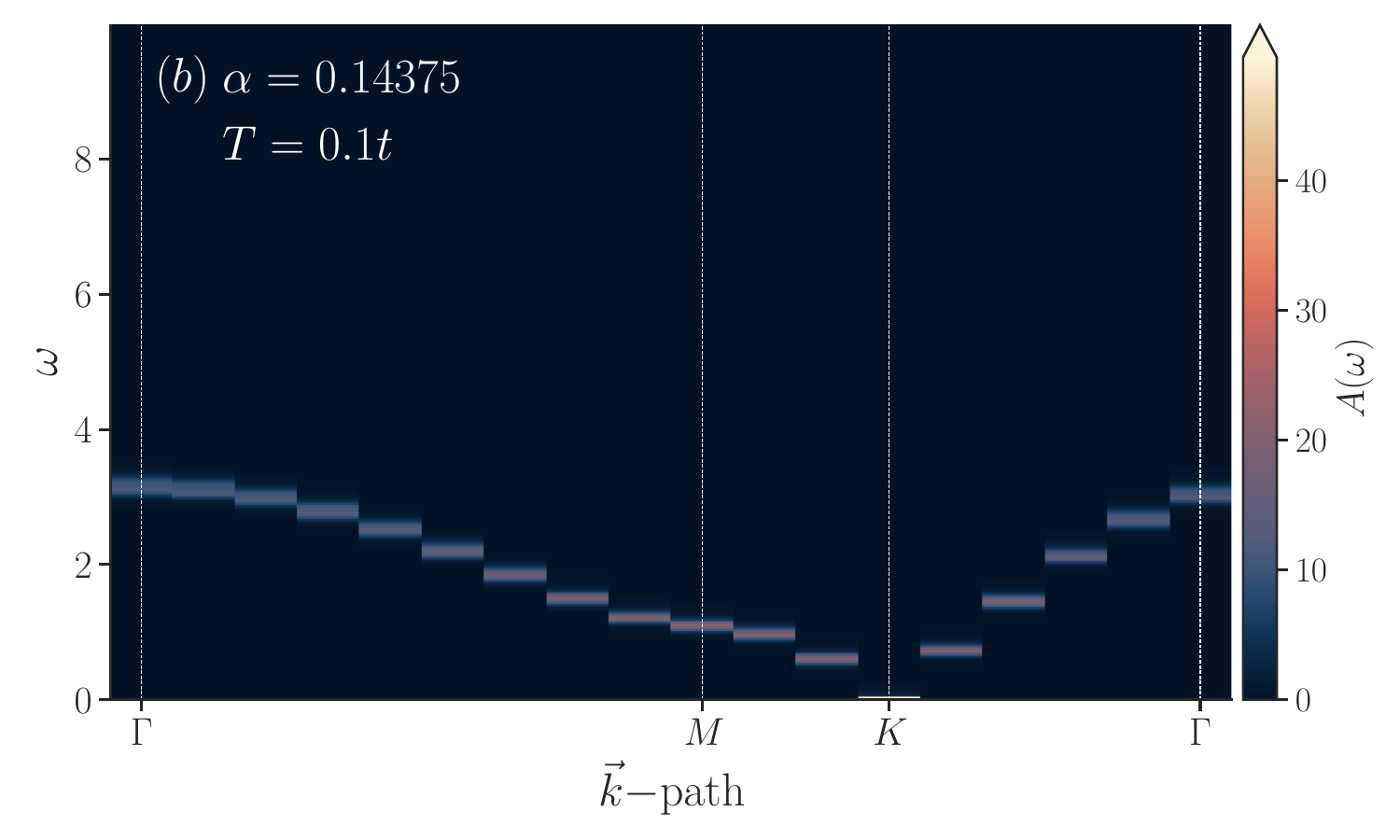}}
     \subfigure     { \label{fig:T0.25_large_alpha_1el}\includegraphics[width=0.35\textwidth]{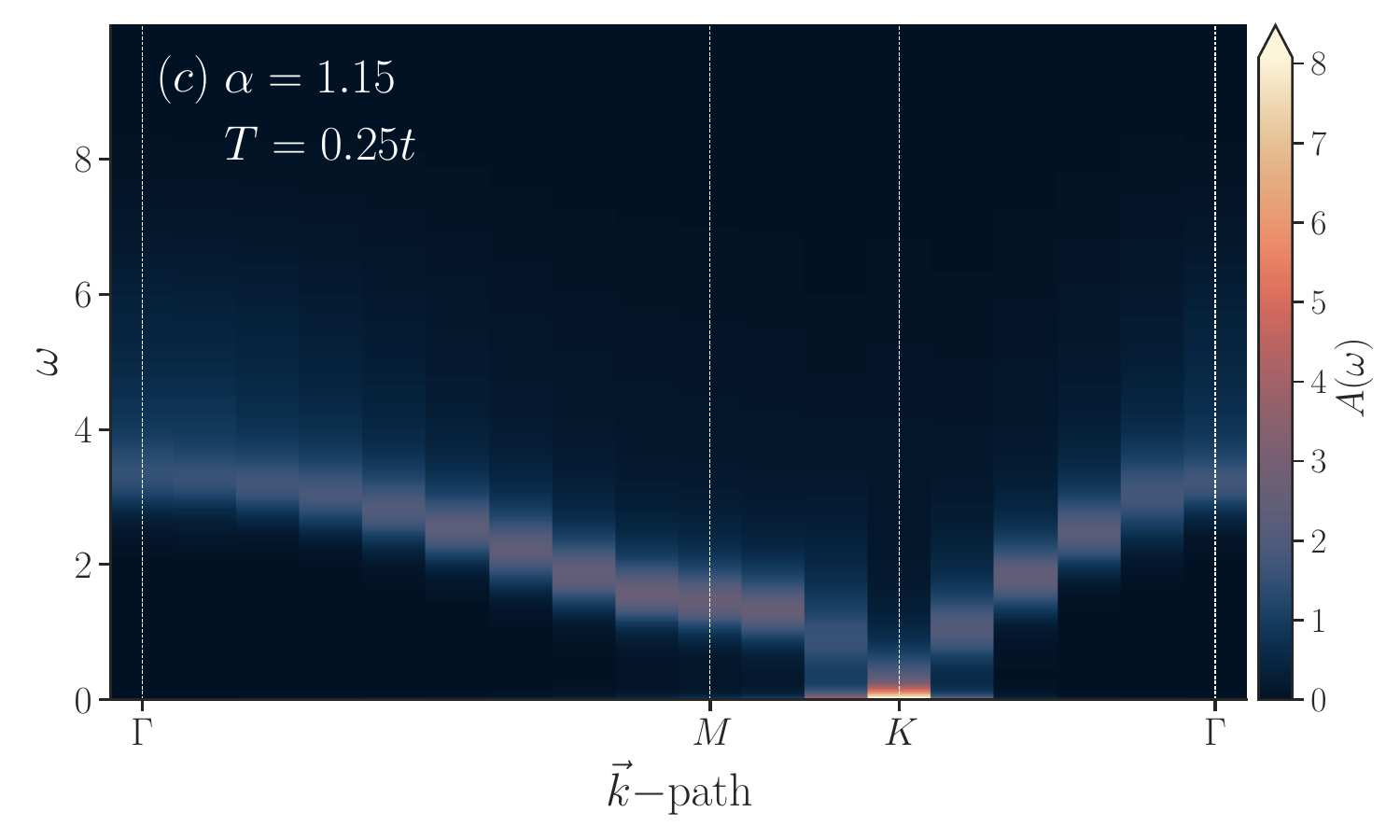}}
      \subfigure     { \label{fig:T0.25_small_alpha_1el}\includegraphics[width=0.35\textwidth]{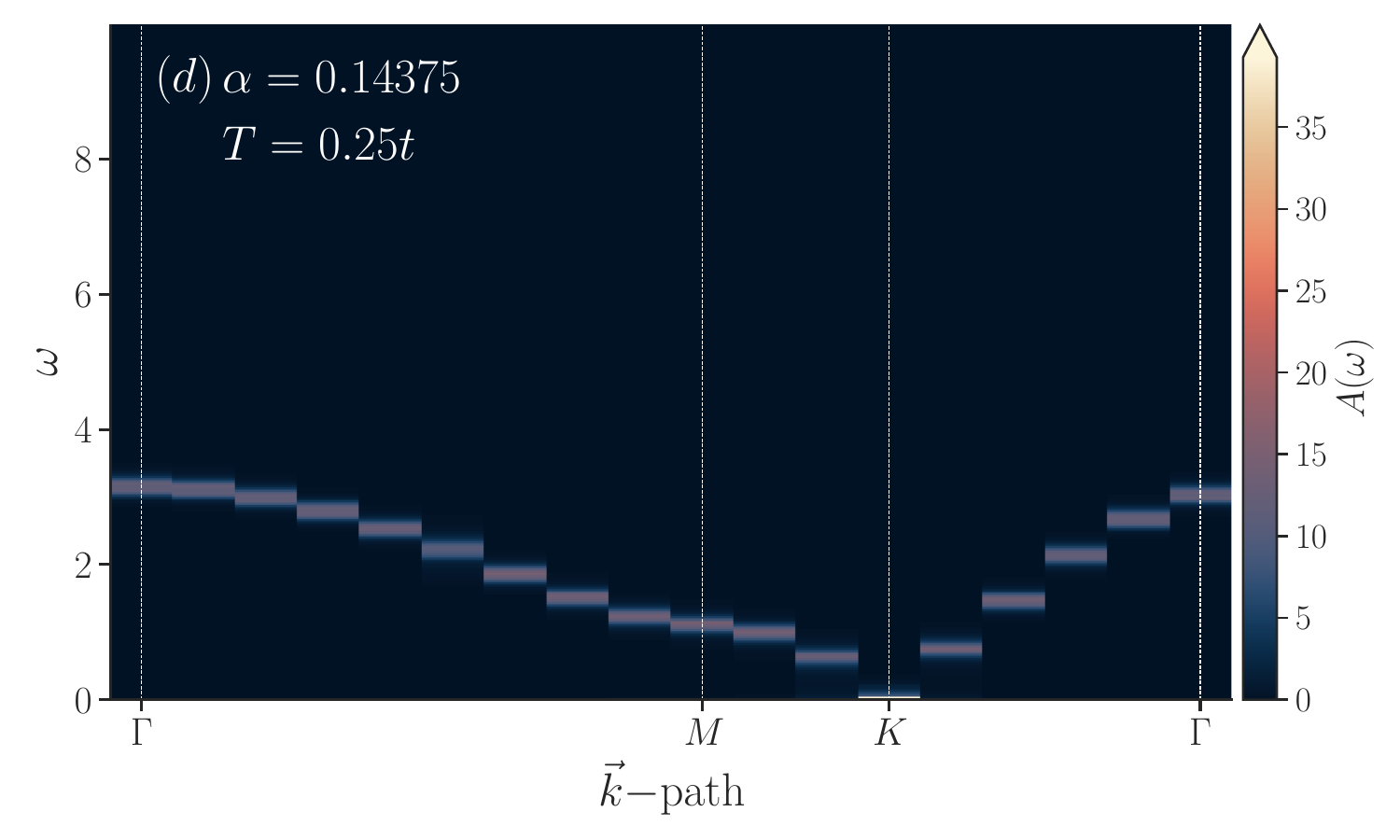}}
         \caption{Single-electron spectral functions for the "Full Coulomb" setup from section \ref{sec:model}. (a) $T=0.1$ and large interaction $\alpha=1.15$; (b) $T=0.1$ and small interaction $\alpha=0.14375$; (c) $T=0.25$ and large interaction $\alpha=1.15$; (d)  $T=0.25$ and small interaction $\alpha=0.14375$. }
    \label{fig:1el_all}
 \end{figure}

  \begin{figure}[]
   \centering
   \subfigure     { \label{fig:T0.1_large_alpha_AFM}\includegraphics[width=0.35\textwidth]{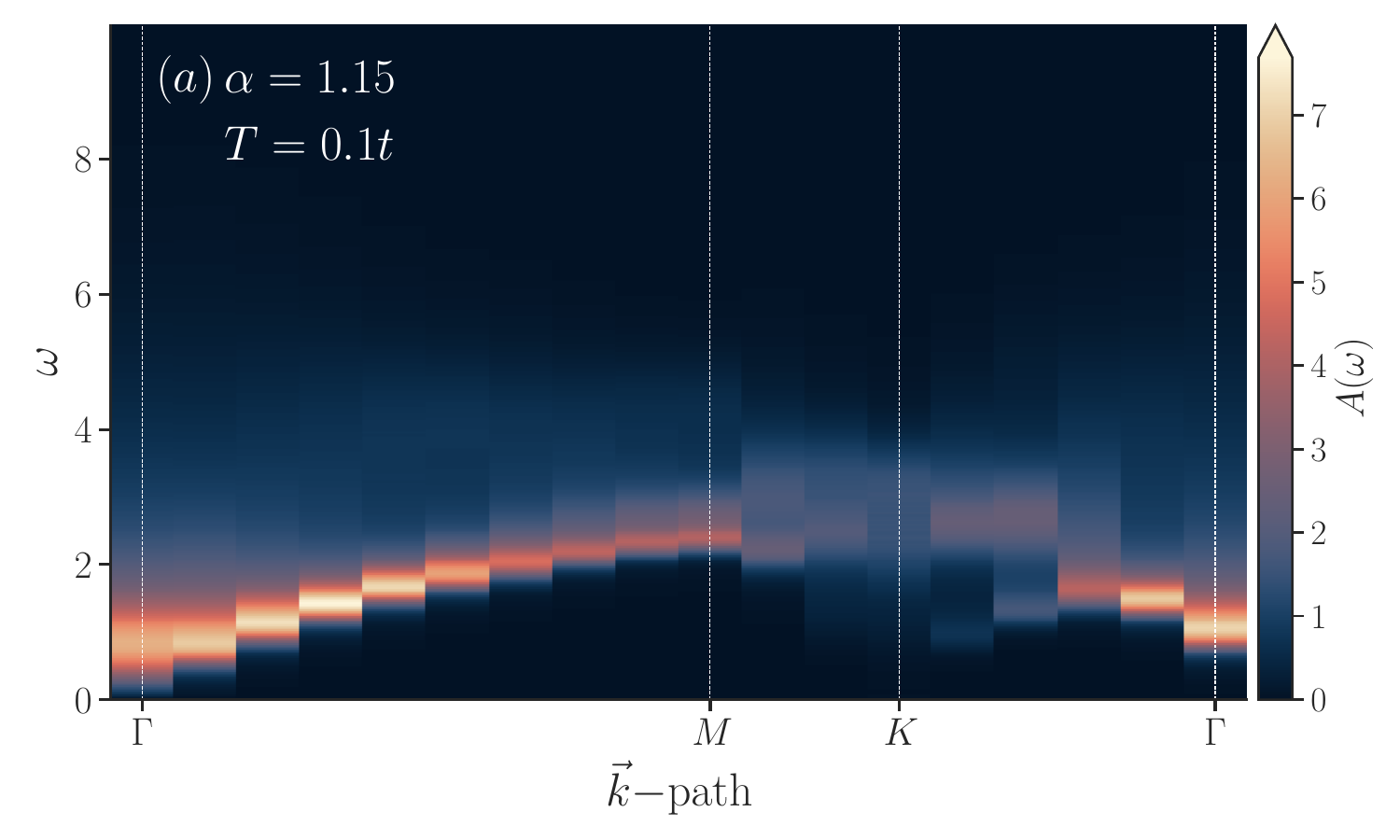}}
    \subfigure     { \label{fig:T0.1_small_alpha_AFM}\includegraphics[width=0.35\textwidth]{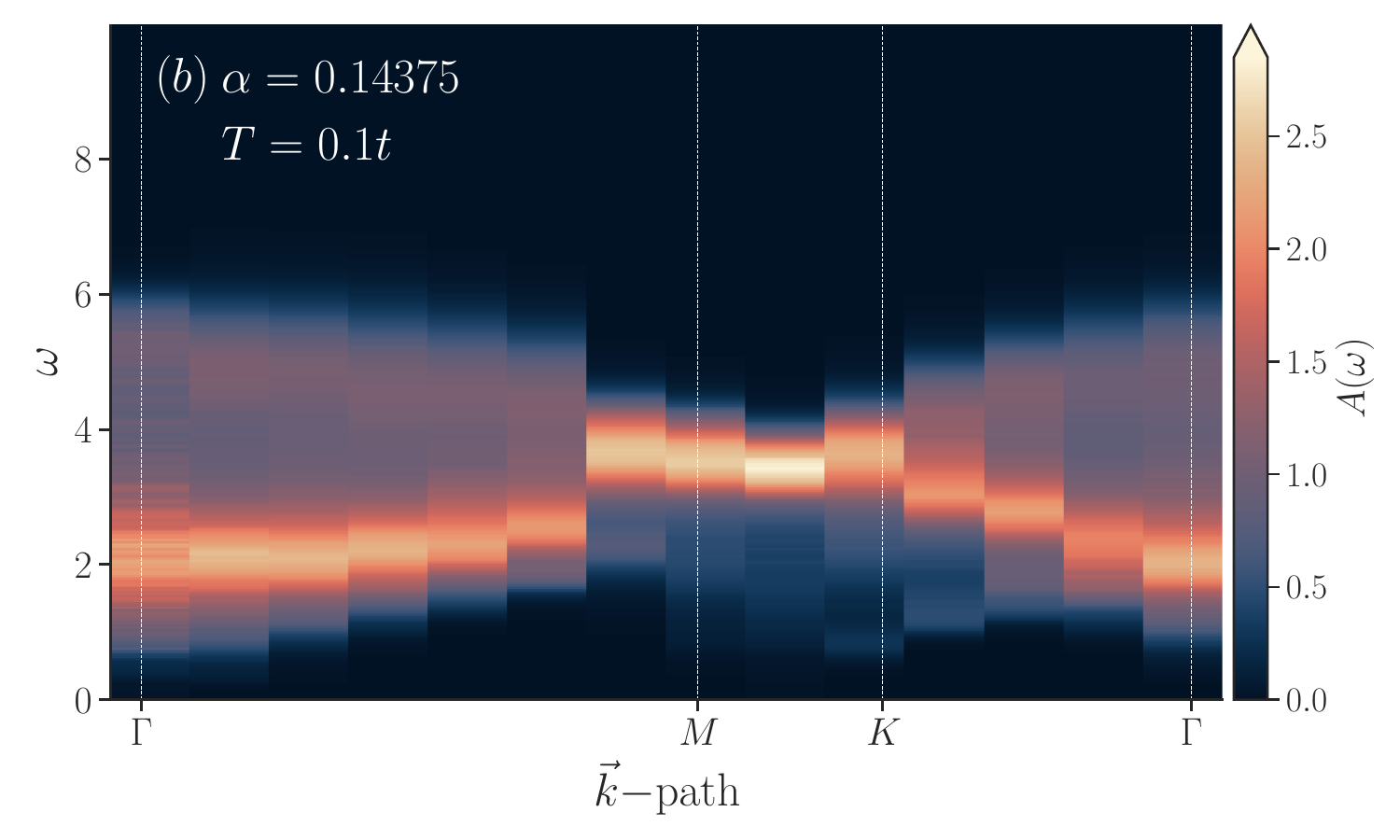}}
     \subfigure     { \label{fig:T0.25_large_alpha_AFM}\includegraphics[width=0.35\textwidth]{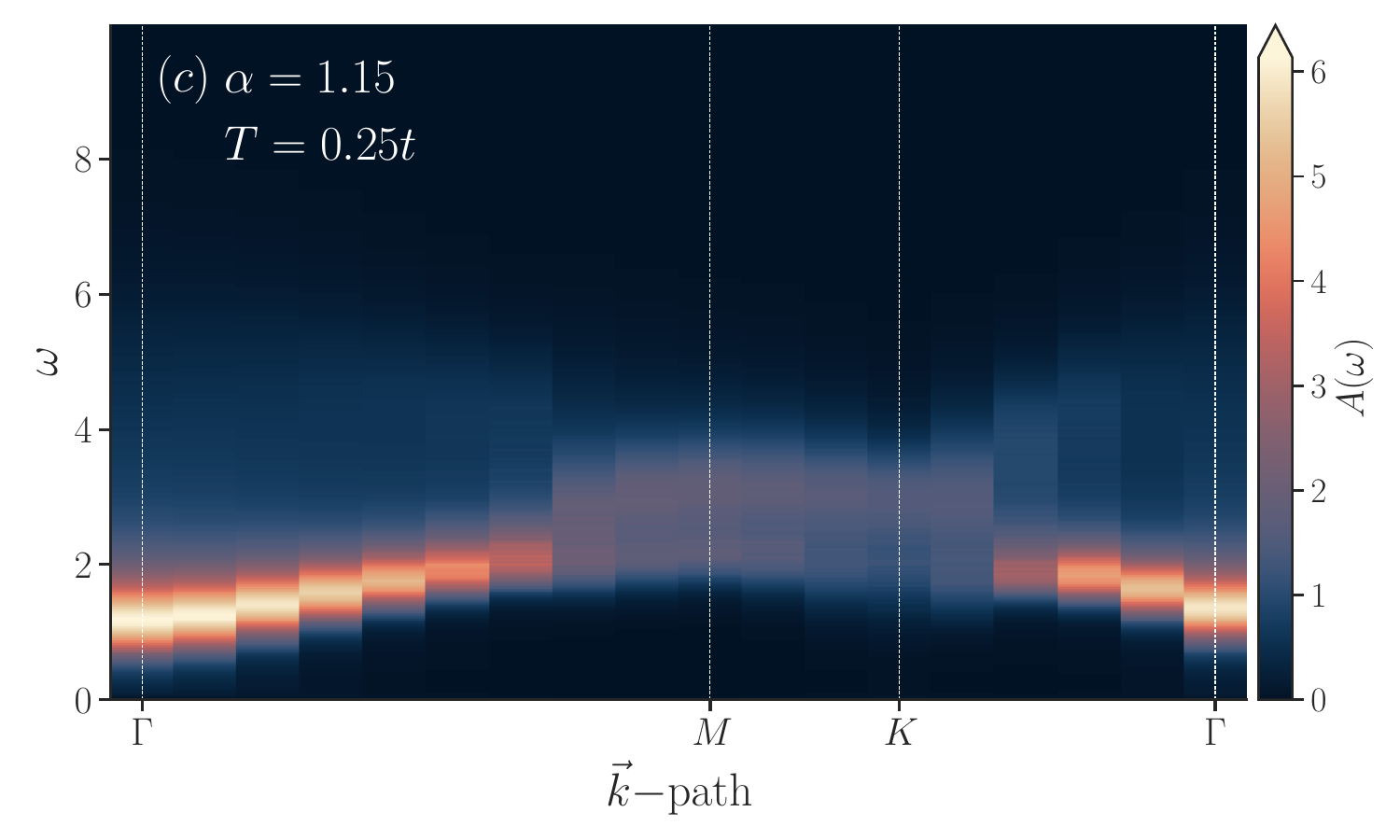}}
      \subfigure     { \label{fig:T0.25_small_alpha_AFM}\includegraphics[width=0.35\textwidth]{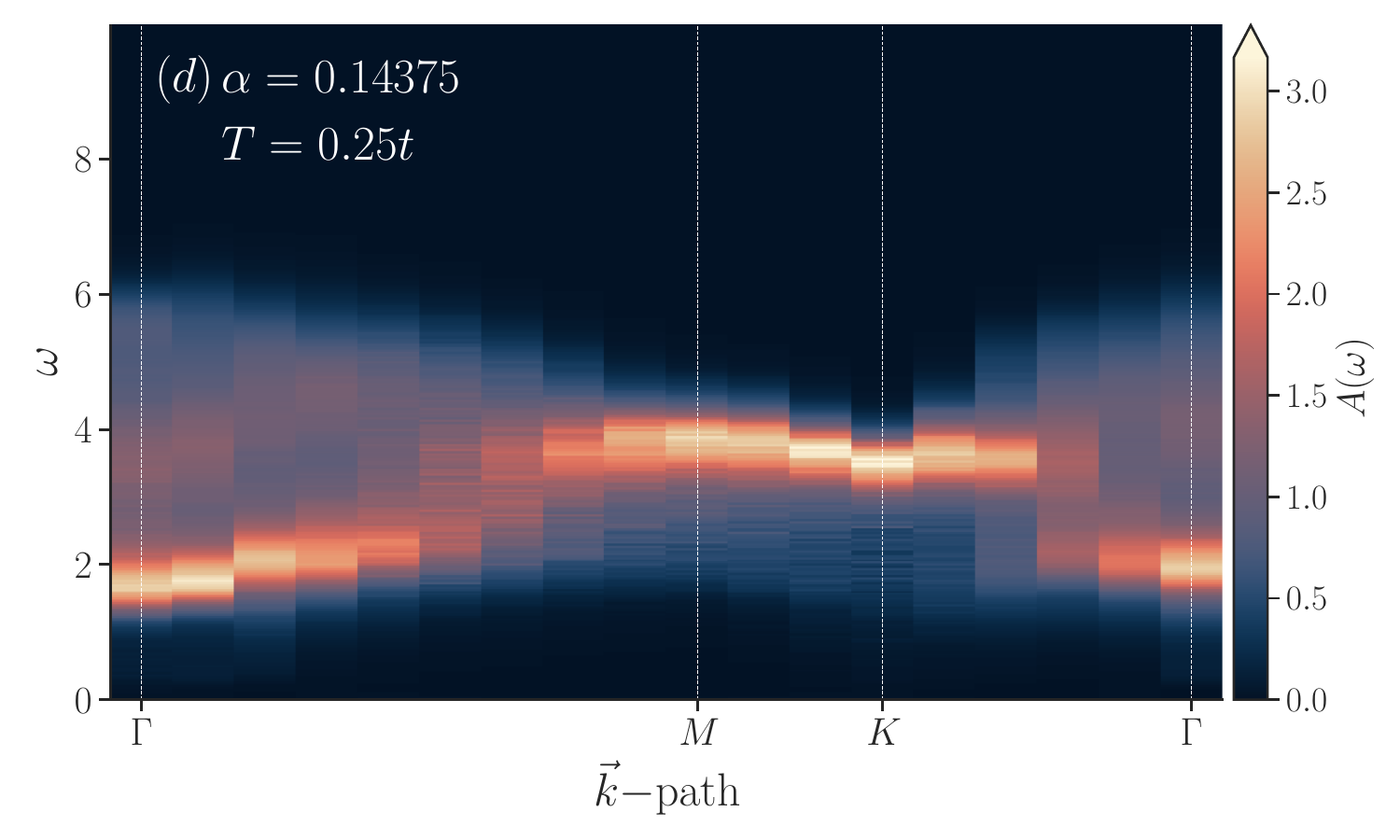}}
         \caption{Spectral functions for AFM spin waves for the "Full Coulomb" setup from section \ref{sec:model}. (a) $T=0.1$ and large interaction $\alpha=1.15$; (b) $T=0.1$ and small interaction $\alpha=0.14375$; (c) $T=0.25$ and small interaction $\alpha1.15$; (d)  $T=0.25$ and small interaction $\alpha=0.14375$.  }
    \label{fig:magnons_all}
 \end{figure}

Let us also mention in passing that, within the Dirac-cone approximation, plasmons are long-lived quasiparticles with square-root energy dispersion 

\begin{equation}
\omega_{\text{pl}}(\vec{k}) = \sqrt{N \alpha v_F  T \ln(4) k} \equiv C\sqrt{k}.
\label{eq:rpa_dispersion}
\end{equation}

However, as it will be evident from subsequent calculations, the square root dependence, even in RPA approximation, only holds up to small frequencies $\omega < \alpha T$. Since most of our QMC results are beyond that limit, instead of comparison with approximate Dirac formula \ref{eq:rpa_dispersion}, we confront QMC results against full numerical lattice RPA. 
 
  \begin{figure}[]
   \centering
   \subfigure     { \label{fig:T0.1_large_alpha_CDW}\includegraphics[width=0.35\textwidth]{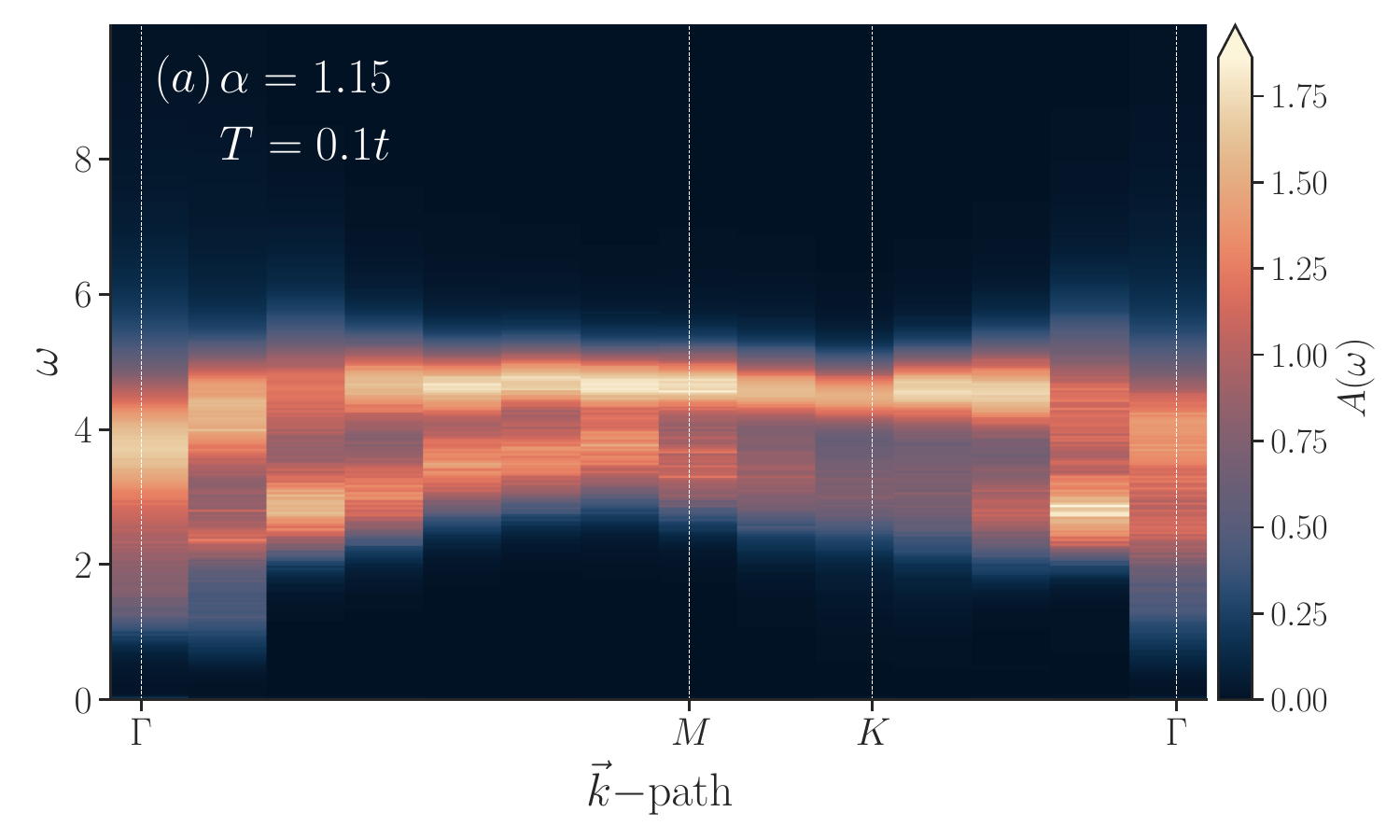}}
    \subfigure     { \label{fig:T0.1_small_alpha_CDW}\includegraphics[width=0.35\textwidth]{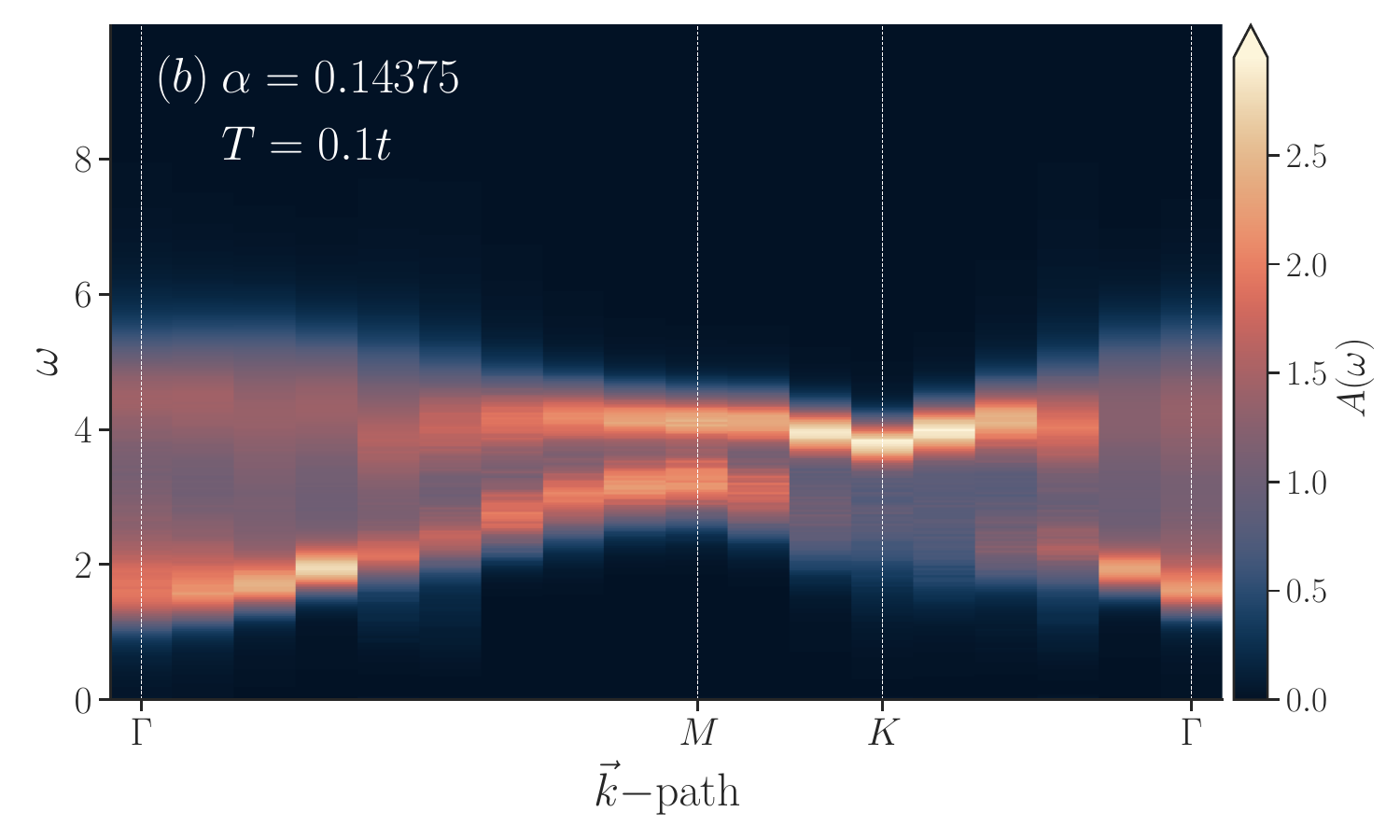}}
     \subfigure     { \label{fig:T0.25_large_alpha_CDW}\includegraphics[width=0.35\textwidth]{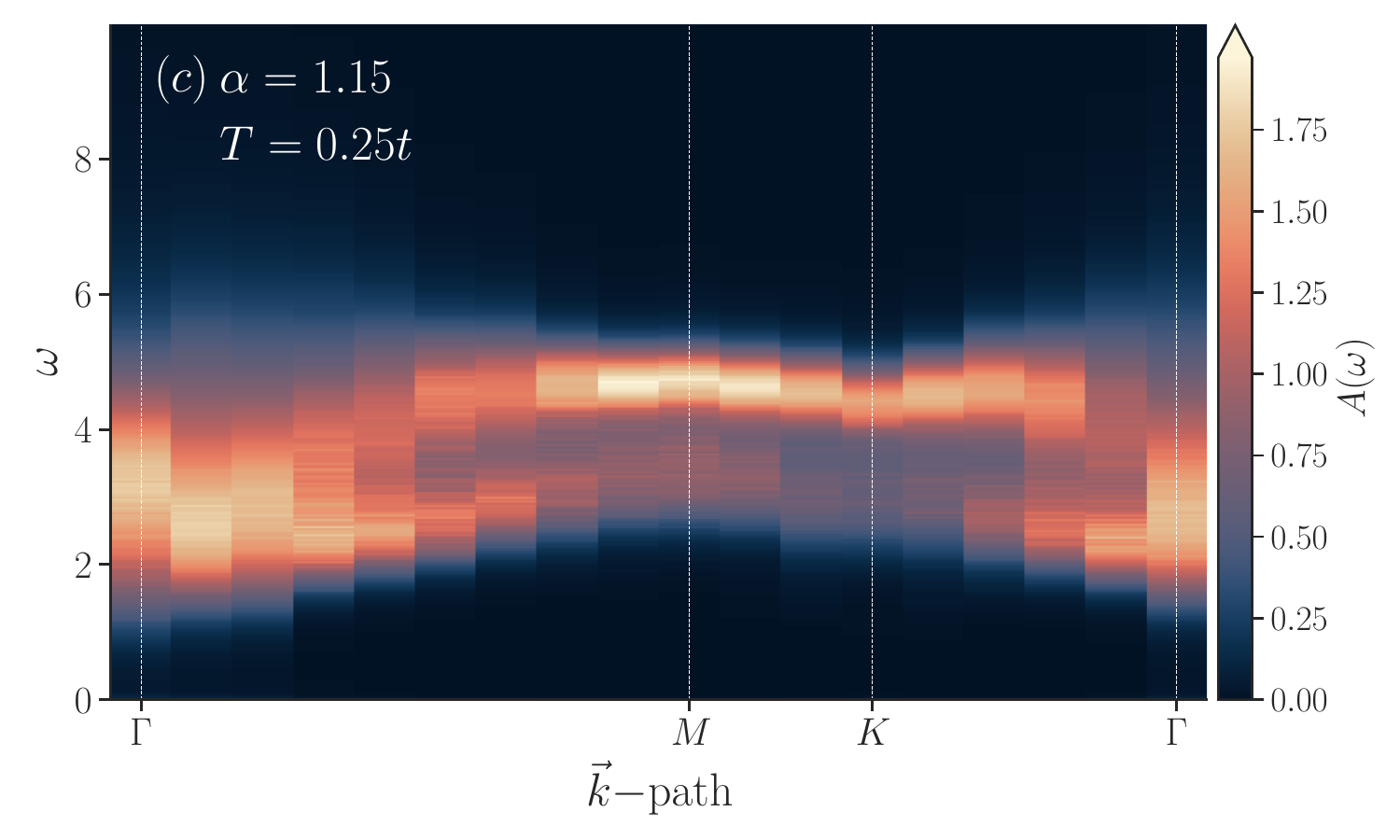}}
      \subfigure     { \label{fig:T0.25_small_alpha_CDW}\includegraphics[width=0.35\textwidth]{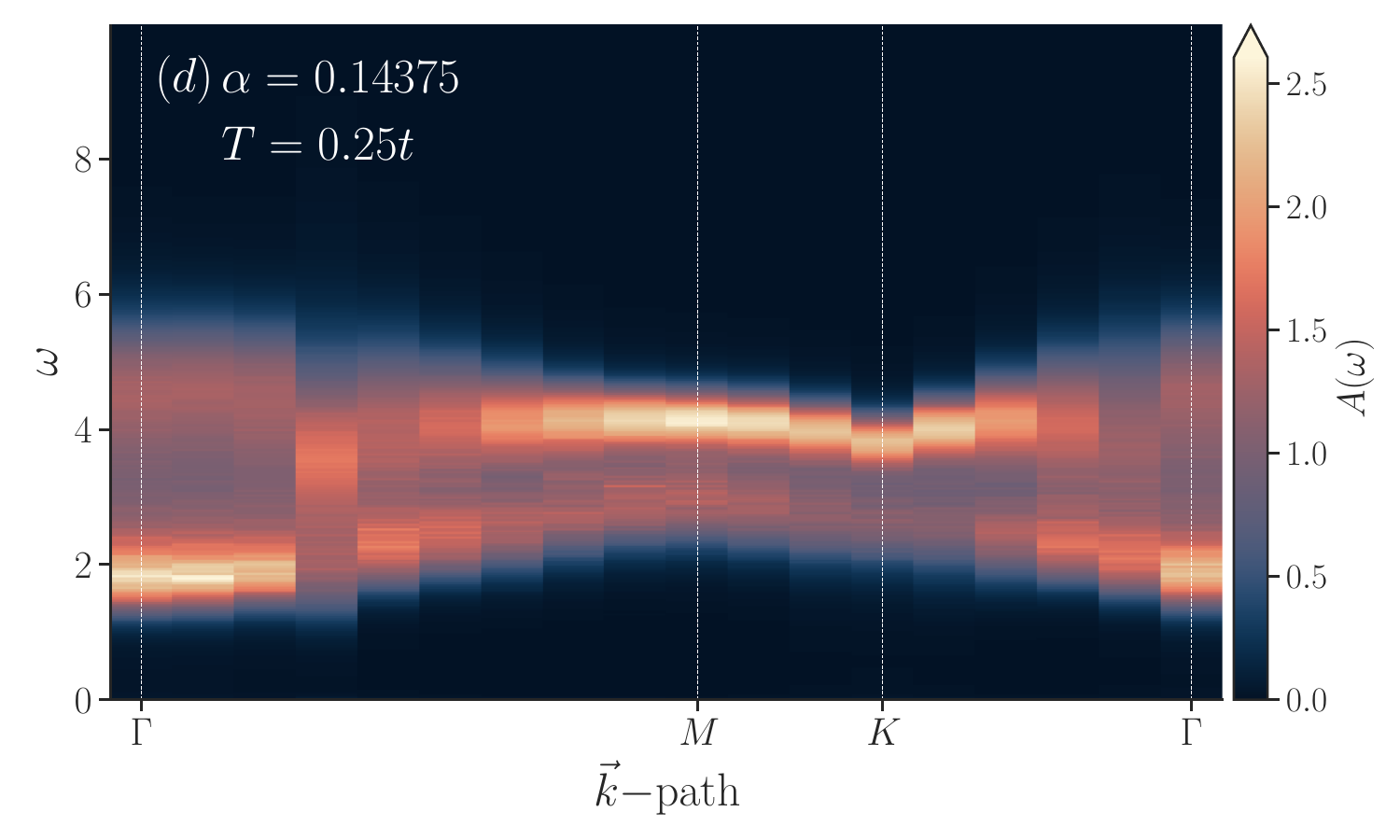}}
         \caption{Spectral functions for CDW excitations for the "Full Coulomb" setup from section \ref{sec:model}. (a) $T=0.1$ and large interaction $\alpha=1.15$; (b) $T=0.1$ and small interaction $\alpha=0.14375$; (c) $T=0.25$ and small interaction $\alpha1.15$; (d)  $T=0.25$ and small interaction $\alpha=0.14375$.  }
    \label{fig:CDW_all}
 \end{figure}

\section{\label{sec:QMC_results}QMC Results}

QMC simulations were performed using the standard BSS-QMC algorithm \cite{Blankenbecler81,White89,Sorella_1992, Assaad08_rev,Ulybyshev13,Hohenadler14}, using the Hamiltonian and long range interactions described in section \ref{sec:model}.
The raw QMC data is organized into 16 batches, each of which is processed independently using stochastic analytical continuation \cite{Beach04a,Sandvik98,SHAO20231}. The final spectral functions are obtained by averaging the resulting spectra over all batches. In addition, we extract the frequency peaks of the plasmonic dispersion for each batch separately, which provides an estimate of the uncertainty in the peak position.

We start from the check that we indeed have gapless electronic excitations with Dirac dispersion. In addition to studying the plasmonic dispersion, we also verify that there are no other gapless collective excitations which can potentially contribute to the transport properties. Thus we compute the following Euclidean time correlation functions in addition to \eqref{eq:corr}:
\begin{eqnarray}
C_O(\vec k, \tau)=\frac{1}{N_{cells}}\sum_{\vec r_1,\vec r_2} e^{-i \vec k (\vec r_1-\vec r_2)}\langle \hat O_{\vec r_1}(0) \hat O_{\vec r_2}^\dag(\tau)\rangle,
\label{eq:corr_add}
\end{eqnarray}
with
\begin{itemize}
    \item  $\hat O_{\vec r}=\hat a_{(\vec r, 0), \uparrow}$ for single-electron excitation. 
      \item $\hat O_{\vec r}=\hat q_{(\vec r, 0)} - \hat q_{(\vec r, 1)}$ for Charge Density Wave (CDW). 
      \item  $\hat O_{\vec r}=\hat S^{(z)}_{(\vec r, 0)} - \hat S^{(z)}_{(\vec r, 1)}$ for Antiferromagnetic (AFM) spin waves, where the spin operator is defined as $\hat S^{(z)}_x = \hat a^\dag_{x, \uparrow} \hat a_{x, \uparrow} - \hat a^\dag_{x, \downarrow} \hat a_{x, \downarrow}$. 
\end{itemize}

\begin{figure}[]
   \centering
   \subfigure     { \label{fig:T0.1_large_alpha_plasmons}\includegraphics[width=0.4\textwidth]{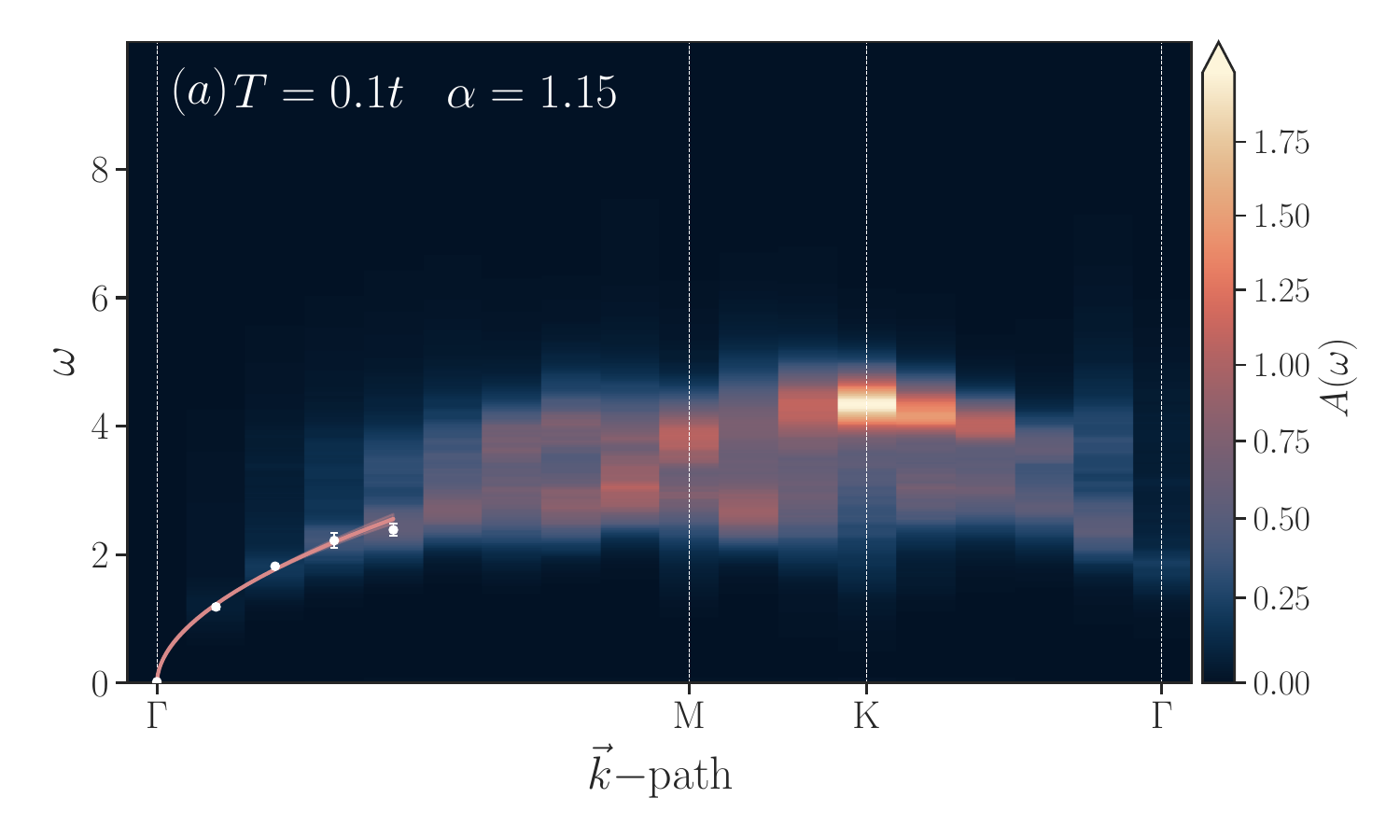}}
    \subfigure     { \label{fig:T0.1_small_alpha_plasmons}\includegraphics[width=0.4\textwidth]{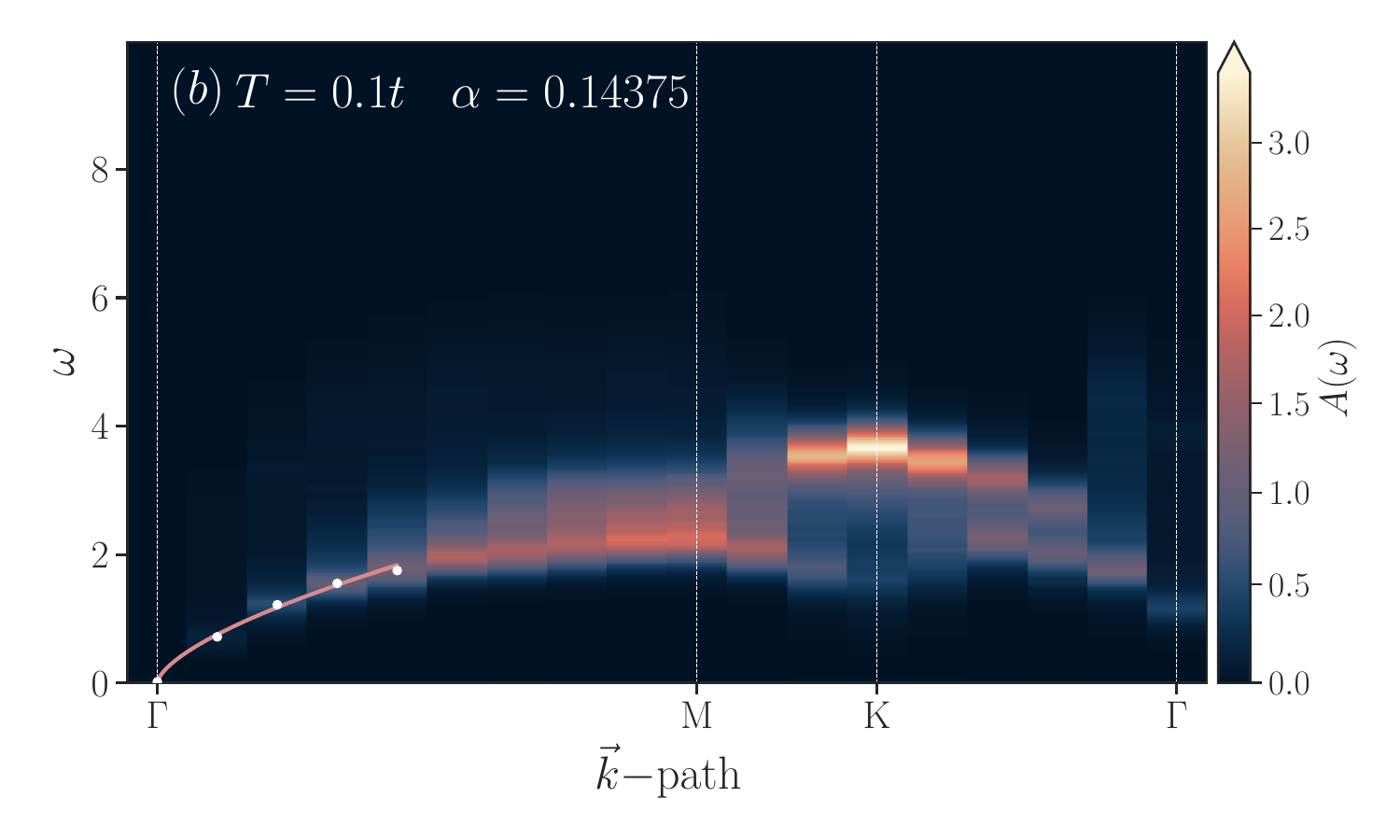}}
     \subfigure     { \label{fig:T0.25_large_alpha_plasmons}\includegraphics[width=0.4\textwidth]{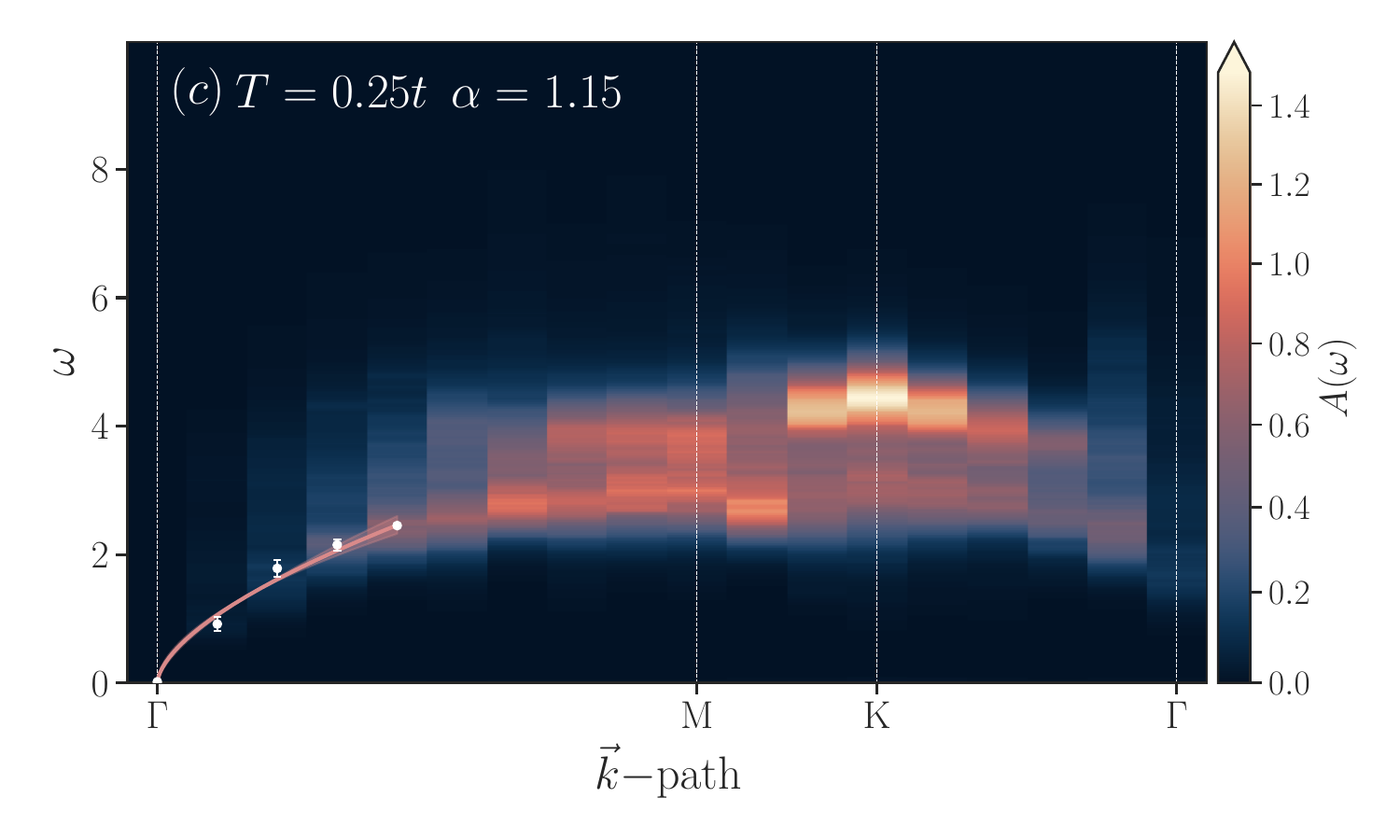}}
      \subfigure     { \label{fig:T0.25_small_alpha_plasmons}\includegraphics[width=0.4\textwidth]{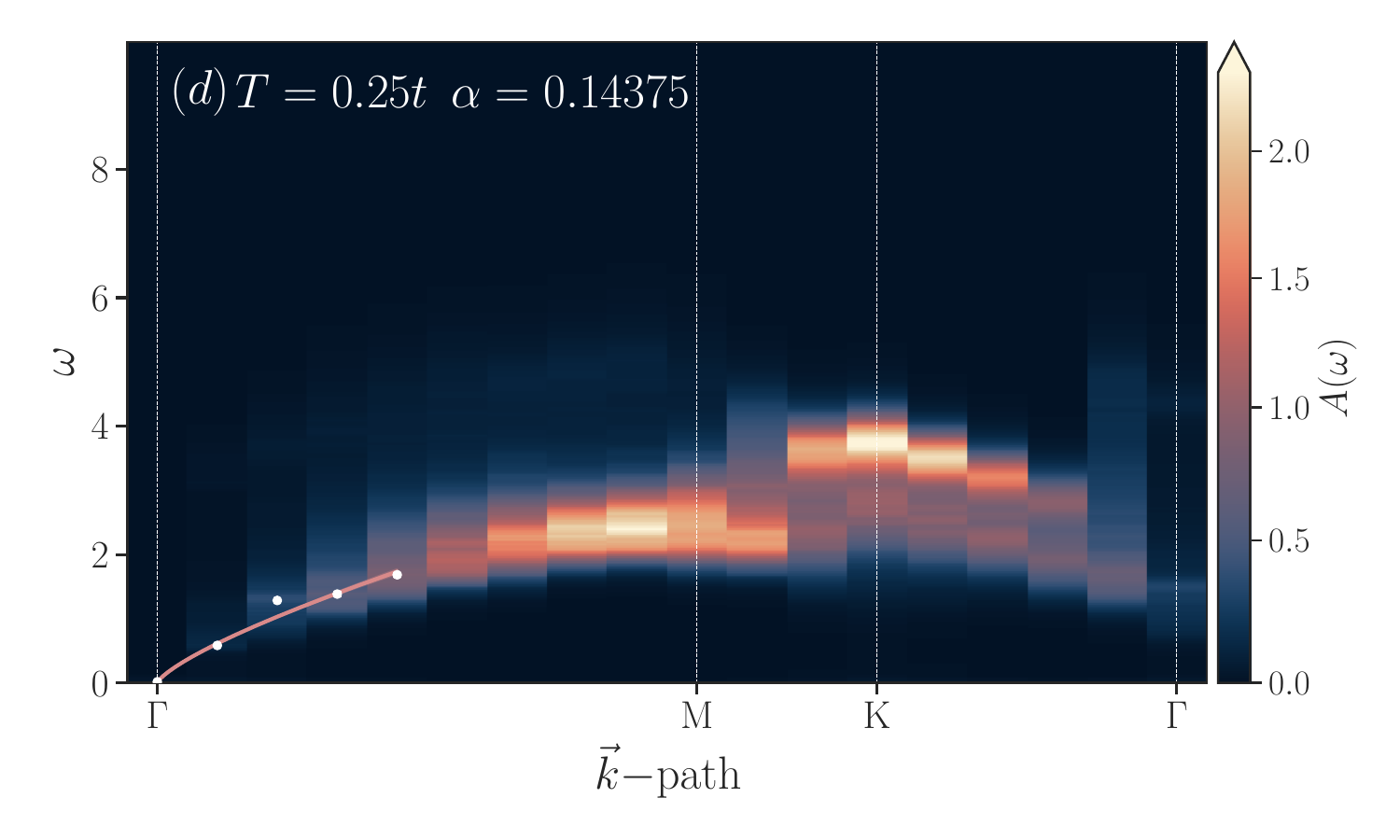}}
         \caption{Spectral functions for plasmons for the "Full Coulomb" setup from section \ref{sec:model}. (a) $T=0.1$ and large interaction $\alpha=1.15$; (b) $T=0.1$ and small interaction $\alpha=0.14375$; (c) $T=0.25$ and large interaction $\alpha=1.15$; (d)  $T=0.25$ and small interaction $\alpha=0.14375$. The dispersions computed at the respective points of the BZ path are included for comparison (see Fig.~\ref{fig:plasmons_all_model_dispersion}).}
    \label{fig:plasmons_all}
 \end{figure}

Figure \ref{fig:1el_all} shows the single particle spectral functions at different temperatures and interaction strength, confirming that our study corresponds to the semimetallic regime, despite the large value of on site interaction $U=4.0t$, which exceeds the critical coupling for the Hubbard model on the hexagonal lattice. However, in our case, the long-range interactions offset the effects of the Hubbard term, and the  electrons are therefore still not gapped. The only noticeable effects of the strong Coulomb interactions are: 1) broadening of the spectral functions around the $\Gamma$-point(Fig.  \ref{fig:T0.1_large_alpha_1el} and \ref{fig:T0.25_large_alpha_1el}); and 2) renormalization of the Fermi velocity in the vicinity of the \K-point, which is especially pronounced at the the largest interaction strength and temperature (Fig. \ref{fig:T0.25_large_alpha_plasmons}). The latter effect of increased  $v_{\mathrm{F}}$ renormalization at higher temperatures is also in agreement with the previously reported results in \cite{Ulybyshev2023}. 

Figure~\ref{fig:magnons_all} shows the spectral functions for AFM spin waves. In agreement with the results for the single-particle spectral functions, our interactions are smaller than the ones needed to drive the system to the AFM Mott insulator, thus the spin waves are gapped. Nevertheless we can see that the gap in the spin waves dispersion is narrowing with the increased interaction strength (e.g. compare Fig. \ref{fig:T0.1_large_alpha_AFM} and Fig. \ref{fig:T0.1_small_alpha_AFM}), potentially leading to gapless spin waves at the phase transition point. 
Finally, Fig.~\ref{fig:CDW_all} confirms that the CDW excitations are always gapped, and therefore do not play a substantial role in the transport properties. \\

 \begin{figure}[]
  \centering
\includegraphics[width=0.45\textwidth , angle=0]{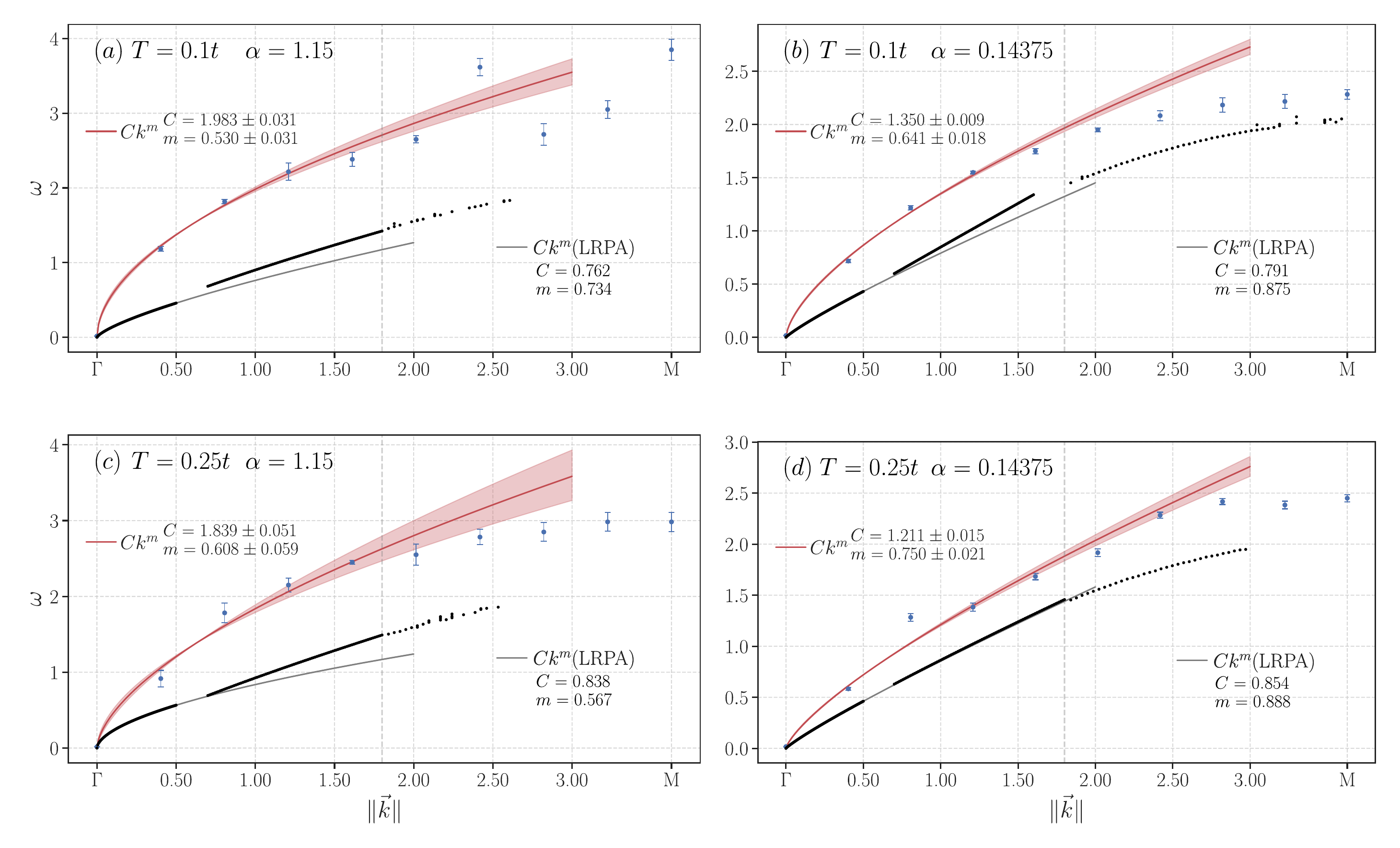}
         \caption{QMC and lattice RPA plasmon dispersions for the full Coulomb setup. In the QMC case, the points are obtained from the peak positions of the spectral function averaged over all data batches, while the uncertainties are estimated from the distribution of peak frequencies extracted from each individual batch.  Momentum $k$ is in the units of $1/a$, where $a$ is the graphene unit cell length and frequencies $\omega$ are in the units of hopping $t = 1$. Fitting with $C k^m$ function is made using points up to $\|\vec{k}\|=1.75$.}
    \label{fig:plasmons_all_model_dispersion}
\end{figure}

 \begin{figure}[]
  \centering
\includegraphics[width=0.45\textwidth , angle=0]{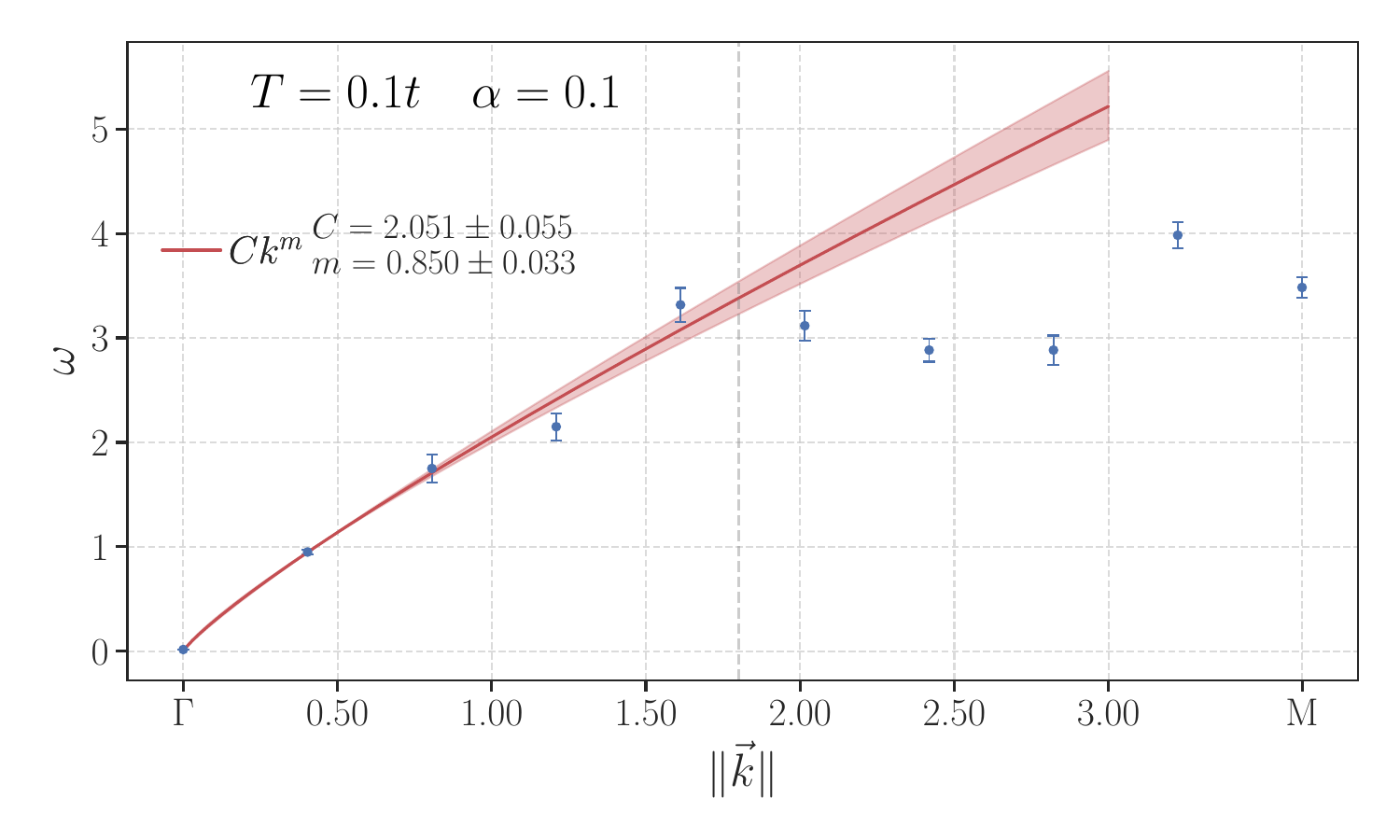}
         \caption{Dispersion relation extracted  from the data of the  "Coulomb $>$ NNN" setup from section \ref{sec:model} where interactions up to next-nearest neighbours are dominating. The dispersion differs strongly from the $\sqrt{k}$ expectation.}
    \label{fig:Screened_Plasmons}
\end{figure}

 \begin{figure}[]
  \centering
\includegraphics[width=0.45\textwidth , angle=0]{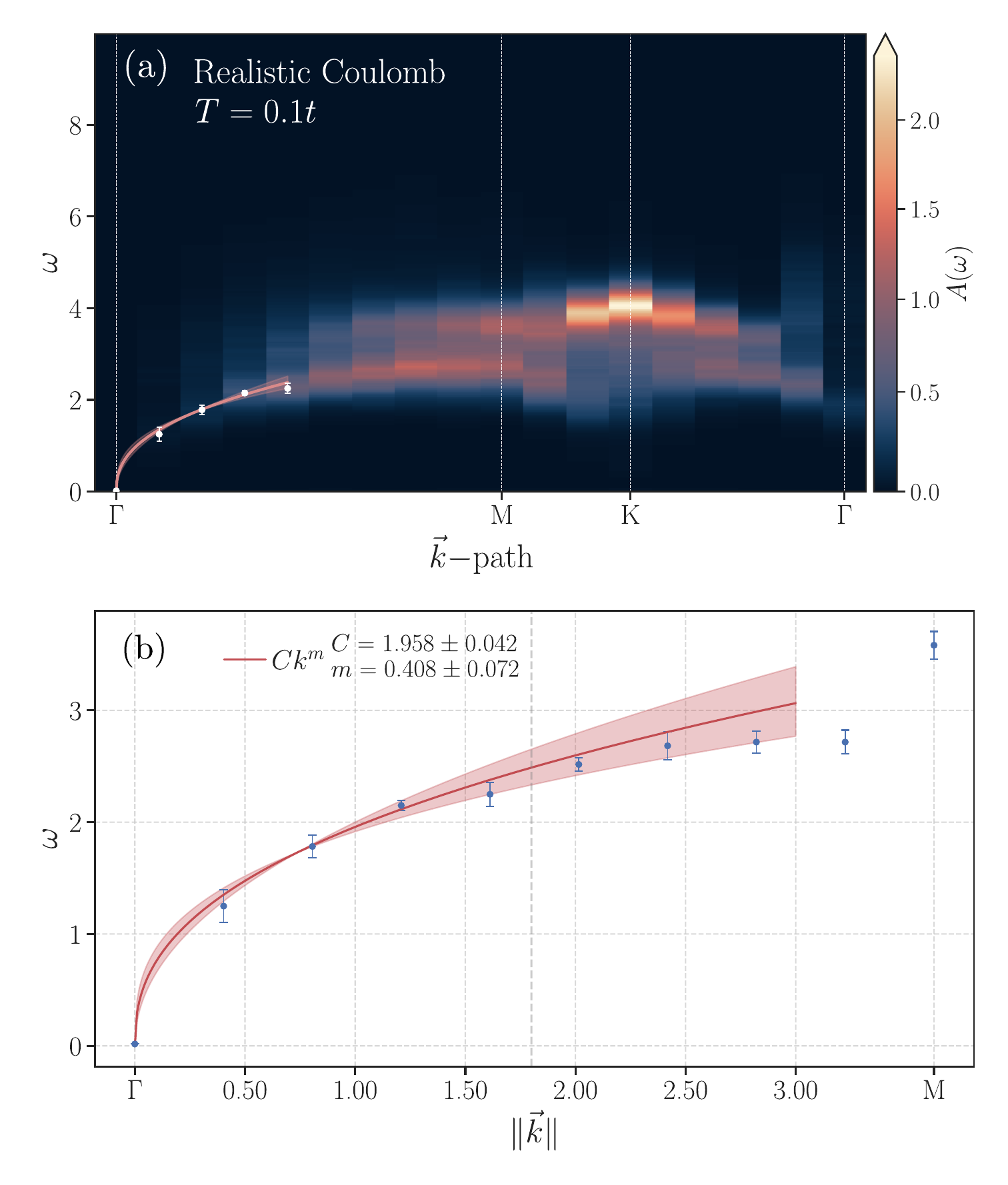}
         \caption{Plasmonic spectral function (a) and extracted dispersion (b) for the setup with realistic Coulomb interactions of Graphene.}
    \label{fig:RealisticGraphene}
\end{figure}

 \begin{figure}[]
  \centering 
\includegraphics[width=0.45\textwidth , angle=0]{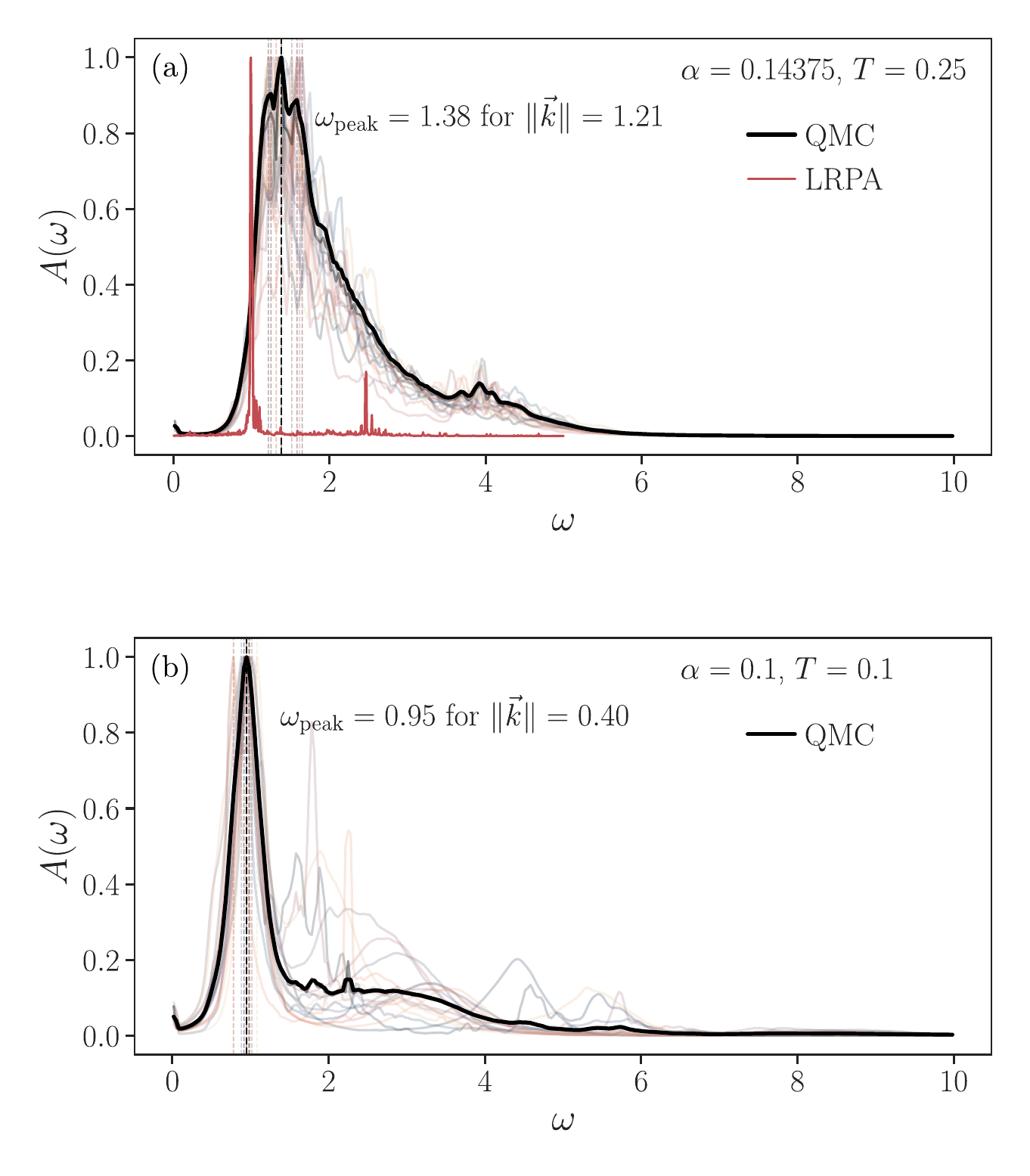}
 \caption{(a) Comparison of the spectral function between the RPA calculation and the QMC simulation for the full Coulomb potential.
(b) A spectral function for the setup with screened Coulomb interactions  "Coulomb $>$ NNN".}
  \label{fig:SpectralFunctionComparison}
\end{figure}

Finally, Figure \ref{fig:plasmons_all} shows the spectral function for plasmons  $\Im C_q(\vec k, \omega)$ (see eq. \ref{eq:GK}) in the full Coulomb interaction setup with two different values of $\alpha$ (c.f. section \ref{sec:model}) and two temperatures. We can already see that the dispersion for plasmons is clearly trending towards zero in the vicinity of the $\Gamma$-point, although with substantial reduction of the spectral weight. \\ Additionally, a secondary plasmonic branch becomes visible in the vicinity of the K‑point but at higher energies. As the interaction strength and temperature increases, this branch propagates deeper into the Brillouin zone and exhibits a pronounced enhancement in spectral weight. The combination of these qualitative features strongly suggests that this branch corresponds to the mode predicted by Levitov and Lewandowski in Ref.~\cite{Levitovplasmon2019}. This mode was also visible as interband term in Fig.~\ref{fig:LRPA} within the RPA approximation, albeit with different quantitative features. 

From the peaks of the spectral functions at different momenta, we extract the (lower) plasmonic energy dispersion, as shown in Figure \ref{fig:plasmons_all_model_dispersion}. The dispersion for the screened Coulomb setup ( "Coulomb $>$ NNN") is obtained in the same manner and presented in Figure \ref{fig:Screened_Plasmons}.\\

Fitting to $C k^m$ reveals that the QMC data cannot be accurately described by either a linear or a square‑root model. Rather, it shows an intermediate scaling form that shifts towards linear or $\sqrt{k}$‑like behavior depending on the interaction strength and on whether the short‑range interaction terms (Fig.~\ref{fig:Screened_Plasmons}) or the long‑range Coulomb interaction (Fig.~\ref{fig:plasmons_all_model_dispersion}) governs the dynamics.\\

For the Full Coulomb setup, we observe that the QMC data converges to the lattice RPA results in the limit of small $\alpha$, with a convergence rate that depends on the temperature. For the strongest interactions, the square-root dispersion is nearly restored, indicating that the scale at which the $\sqrt{k}$ behaviour is valid is set by the product $\alpha T$, visible in both  QMC and LRPA results (Tab.~\ref{tab:m_coef_table}).

The mismatch of the coefficients $C$ and $m$ from LRPA and QMC, reported in Tables ~\ref{tab:m_coef_table} and ~\ref{tab:C_Coeff_table}, can not be attributed to a finite Brillouin zone, since both computations are carried out for the same lattices. Instead, it should be attributed to the the inherent presence of higher order scattering processes in the QMC computations. Some role is also played by the additional Hubbard on-site interaction, not taken into account in the LRPA computations. The mismatch between Dirac RPA and LRPA, which is (at fixed momenta) more pronounced for weaker interactions and thus smaller $\alpha T$ can however be attributed to the finite Brillouin zone of the hexagonal lattice which can alter the RPA-level diagrams in comparison to the Dirac cone approximation \cite{Ulybyshev2023}.

Since both "Full Coulomb" and "Coulomb $>$ NNN" setups are made somewhat artificial in order to highlight differences between short-range interaction-dominant and long-range interaction-dominant regimes, we have checked that the "Realistic Coulomb" setup features qualitatively similar plasmonic excitations. Corresponding results of the plasmonic spectral function and  dispersion are shown in Figure~\ref{fig:RealisticGraphene}. We see almost exactly the same spectral functions as for the strong interaction case ("Full Coulomb" setup). Due to the effective $\alpha\sim 1$, the fitted dipsersion exhibits a square-root dependence up to higher frequencies. In addition, we also observe the second "interband" branch of plasmons once we approach the M-point in the BZ.  

\begin{table}[htbp]
\centering
\begin{tabular}{l|cccc}
\hline\hline
\multirow{2}{*}{\Large $m$} & \multicolumn{2}{c}{$\alpha = 1.15$} & \multicolumn{2}{c}{$\alpha = 0.14375$} \\
\cline{2-3} \cline{4-5}
 & $T=0.1t$ & $T=0.25t$ & $T=0.1t$ & $T=0.25t$ \\
\hline
QMC  & $0.530 \pm 0.031$ & $0.608 \pm 0.051$ & $0.641 \pm 0.018$ & $0.854 \pm 0.021$ \\
LRPA & $0.734$ & $0.567$ & $0.875$ & $0.888$ \\
\hline\hline
\end{tabular}
\caption{Comparison between the exponent $m$ in the $C K^m$ fit of the plasmon dispersion obtained from QMC calculations and LRPA
    (see Fig.~\ref{fig:plasmons_all_model_dispersion})}
\label{tab:m_coef_table}
\end{table}

\begin{table}[htbp]
\centering
\begin{tabular}{l|cccc}
\hline\hline
\multirow{2}{*}{\Large $C$} & \multicolumn{2}{c}{$\alpha = 1.15$} & \multicolumn{2}{c}{$\alpha = 0.14375$} \\
\cline{2-3} \cline{4-5}
 & $T=0.1t$ & $T=0.25t$ & $T=0.1t$ & $T=0.25t$ \\
\hline
QMC  & $1.983 \pm 0.031$ & $1.839 \pm 0.051$ & $1.350 \pm 0.009$ & $1.211 \pm 0.015$ \\
LRPA & $0.762$ & $0.838$ & $0.641$ & $0.750$ \\
RPA  & $0.56$ & $0.89$ & $0.19$ & $0.32$ \\
\hline\hline
\end{tabular}
\caption{Comparison of the coefficient $C$ between the LRPA and QMC fits of the form $C K^{m}$ (see Fig.~\ref{fig:plasmons_all_model_dispersion}), and the RPA result of the form $C\sqrt{K}$. }
\label{tab:C_Coeff_table}
\end{table}

The role of higher order scattering processes becomes also visible in the comparison of the spectral functions obtained from the LRPA calculation and the QMC simulation in the weak-interaction regime at low temperature (Figure~\ref{fig:SpectralFunctionComparison}). We find that LRPA predicts plasmons to be more stable than those observed in the QMC simulations. This difference arises because QMC accounts for a broader range of scattering processes than the LRPA approximation. In LRPA, plasmons acquire a finite lifetime solely due to Landau damping. In contrast, the QMC simulations indicate that additional processes play an important role. Because both calculations use finite lattices with finite Brillouin zones, the effect cannot be attributed to finite‑size limitations.

 \begin{figure}[]
  \centering 
\includegraphics[width=0.4\textwidth , angle=0]{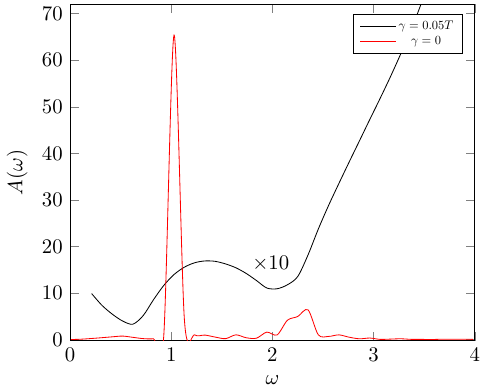}
        \caption{Comparison of the plasmon spectral functions for graphene at temperature $T=t$ and momentum $k=0.4371a_{CC}$ for free electrons ($\gamma=0$) and dressed electrons with $\gamma=0.05t$. As for a comparison, graphene encapsulated in hBN has transport lifetime around $0.2T$ \cite{Gallagher2019} in the hydrodynamic regime. }
  \label{fig:spectralfunctionwithinteraction}
\end{figure}

To illustrate the influence of electron-electron scattering on the plasmon lifetime, we can calculate the polarization function using a dressed electron Green function with a constant broadening $\gamma$, i.e., the retarded Green function is given by
\begin{equation}
g^R_\lambda(\vec{k},\omega) = \frac{1}{\omega+i\gamma-\epsilon_\lambda(\vec{k})},
\end{equation}
and advanced Green function is given by $g^A(\vec{k},\omega) = (g^R(\vec{k},\omega))^*$.
The Lindhard formula is generalized to 
\begin{eqnarray}
&&\Pi(\vec{k},\omega)=-\frac{iN}{2} \sum_{\lambda,\lambda'=\pm}  \int \frac{d\vec{q}}{(2\pi)^2}\frac{d\nu}{2\pi}\mathcal{F}_{\lambda\lambda'}(\vec{k},\vec{q}) \nonumber \\ && \left( g^R(\vec{k}+\vec{q},\omega+\nu)g^K(\vec{q},\nu) + g^K(\vec{k}+\vec{q},\omega+\nu)g^A(\vec{q},\nu)\right). \nonumber\\
\end{eqnarray}
Here $g^K(\vec{q},\nu) = 2i\Im g^R(\vec{q},\nu) \tanh(\nu/2T)$. This formula is derived from the Keldysh QFT \cite{kamenev2011} and in the limit $\gamma \rightarrow 0$ it reduces to the Lindhard formula Eq.~\eqref{eq:Lindhard}. As shown in Fig.~\ref{fig:spectralfunctionwithinteraction}, we find that the broadening in the electron spectral function enlarges the particle-hole continuum and significantly reduces the plasmon stability. We also observe that the plasmon frequency peak is shifted towards high energy. 

In order to support our claim that the decreasing electron quasiparticle lifetime is an important reason for the decrease of the plasmon lifetime, we have also performed simulations away of half filling, in the Fermi liquid regime. Unfortunately, QMC simulations of graphene with unscreened long range Coulomb interactions are not possible at finite doping due to the sign problem. For this reason, we have chosen another model, which is known to exhibit only a mild sign problem but still retains the Fermi liquid state. Namely, we simulated a square lattice Hubbard model in the overdoped regime, where the Fermi the surface is much smaller than the BZ. 

\begin{figure}[]
  \centering 
\includegraphics[width=0.45\textwidth , angle=0]{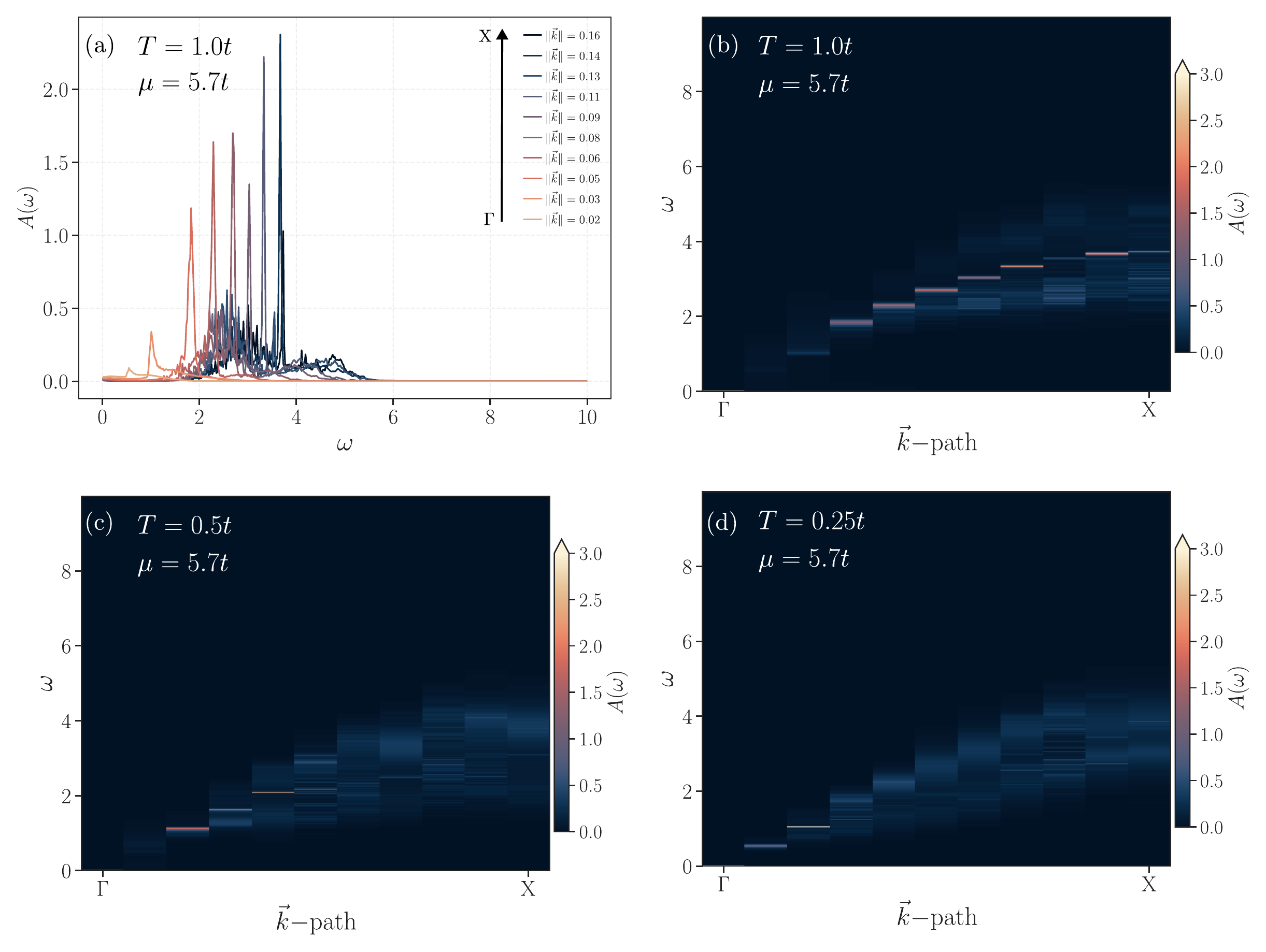}
 \caption{Plasmonic spectral functions for a Fermi-Liquid simulated on a square lattice with Hubbard interaction $U = 6.0t$, away from half-filling.}
  \label{fig:FermiLiquidPlasmons}
\end{figure}

In the Fermi liquid regime, the single-electron quasiparticle lifetime scales as $1/\varepsilon^2$, where $\varepsilon$ is the energy above the Fermi level. If we compare it with the Dirac liquid scaling of $1/\varepsilon$, it is evident that this difference makes electrons in the Fermi liquid regime more long-living in the limit of small energies. The effect of this difference on the plasmonic spectral functions is checked in Figure \ref{fig:FermiLiquidPlasmons}. Indeed we see that plasmons are now very sharp, long lived quasiparticles, and that their widths becomes comparable to the one from RPA in Fig.~\ref{fig:SpectralFunctionComparison} (a). This sharpening  cannot be attributed to the purely short‑range Hubbard interactions, since the comparable screened‑interaction setup (“Coulomb $>$ NNN”) still yields broad spectral functions (see Fig.~\ref{fig:SpectralFunctionComparison} (b)). We therefore conclude that the short plasmonic lifetime is indeed connected to the relatively shorter single-electron quasiparticle lifetime in the Dirac liquid regime at half-filling. 

Finally, wee see that the plasmonic band with sharp quasiparticle peaks is confined to a small region around the $\Gamma$-point, with an extent in momentum-space inversely proportional to temperature, again in agreement with RPA.  Although a fuller investigation of the Fermi‑liquid regime is beyond the scope of this paper, the results shown here already reinforce our claim that our QMC results converge to RPA away from half-filling and motivate a future exploration of this regime.

At last, we should also note that inelastic plasmon–plasmon scattering processes occurring at higher orders of perturbation theory can be responsible for the decrease of the plasmonic lifetime
and investigating these effects constitutes another promising direction for future research.

\section*{\label{sec:conclusions}Conclusions}

We studied the behavior of plasmons in half-filled, free-standing graphene using unbiased QMC calculations and compared the results with the lattice Random Phase Approximation. We confirm the existence of well-defined plasmon resonance peaks around the $\Gamma$-point, and the existence of an additional higher energy plasmonic branch stemming from interband transitions appearing away from the $\Gamma$-point. 

We have determined that if the RPA calculations are performed beyond the Dirac cone approximation, the plasmon dispersion on the lattice follows the standard square root dependency only up the energies of $\omega \approx \alpha T$. Once we are beyond this regime, the LRPA results actually show a linear dispersion. QMC results can be generally fitted with a power law function $C k^m$. At large temperature and large interaction strength, this power law converges to a square-root result due to expanding the $\alpha T$ scale. However, the coefficients in front of the square-root deviates from the ones predicted by LRPA by a factor of two due higher-order perturbative corrections. At small interaction strength we observe convergence of QMC and LRPA results towards a single curve. 

If the Coulomb tail is reduced and the short-range interactions (on-site and nearest-neighbor) dominate, the plasmons are transformed into “first-sound” modes with almost linear dispersion relation in the vicinity of the $\Gamma$-point. 
One may then ask how plasmons can exist at all when only local interactions are present. The answer is that, when both on-site and nearest-neighbor interactions are included, loop corrections generate a renormalized interaction matrix that extends to all distances, thereby effectively producing a long-range interaction. The nearest-neighbor interaction terms are essential for this mechanism: the on-site interaction alone couples only spin-up and spin-down electrons, whereas the nearest-neighbor interaction generates all possible spin combinations. 

In general, these results demonstrate that plasmons are highly sensitive to the details of Coulomb screening and can undergo a crossover into a ``first-sound'' mode with linear dispersion when the long-range part of the electron-electron interaction is screened, while the interactions at distances shorter than the next-to-nearest-neighbor spacing remain unaffected.


It turns out that the plasmons calculated from LRPA are more stable than the ones in our QMC simulations because of the infinite single-electron  quasiparticle lifetime assumed within the LRPA calculations. Once a finite lifetime is artificially inserted in the fermionic propagator, we also get a broadening of the plasmonic excitations, yielding a better agreement with the QMC results. In addition, QMC shows much sharper plasmons in the Fermi liquid regime, where single electrons have a much larger lifetime in the vicinity of the Fermi point than in the Dirac liquid. Another effect not captured by standard RPA calculations is the possibility of inelastic multi-plasmon scattering processes. They appear at higher orders of perturbation theory and can significantly reduce the plasmon lifetime below the one defined by the Landau damping process taken into account at the RPA level. This effect is visible in the spectral functions obtained from QMC simulations, since higher-order interactions are included automatically. 

Experimentally, the reported results (mainly the broadening of the plasmonic resonance and noticeable shift of the plasmonic dispersion upwards in comparison with LRPA prediction especially in case of free-standing graphene) should be noticeable in experiments using energy loss spectroscopy, similar to the ones reported in some experiments in doped graphene \cite{PhysRevB.83.161403, PhysRevB.91.045418, PhysRevB.78.201403}. Since performing similar measurements exactly at half filling is experimentally challenging, we hope that our results, which show that strong correlation effects and deviations from RPA predictions are much more pronounced at half filling, will motivate further experimental efforts in this direction.

\begin{acknowledgments}
MU and AR thank Fakher Assaad for valuable discussions and his support of this project. MU and AR are supported by the DFG through the grant No.\ AS 120/19-1 (Project No.\ 530989922). The project was also partially supported by the W\"urzburg-Dresden Cluster of Excellence on Complexity and Topology in Quantum Matter - \textit{ct.qmat} (EXC 2147, Project No.\ 390858490). 
KP acknowledges former collaborations and discussions with Lars Fritz and Henk Stoof. KP was supported by the
Faculty of Science, Mahidol University. The computational calculations were carried out using JUWELS ~\cite{JUWELS} system (JSC) and the HPC resources provided by the Erlangen National High Performance Computing Center (NHR@FAU) of the Friedrich-Alexander-Universit\"at Erlangen-N\"urnberg (FAU) under the NHR project b133ae. NHR funding is provided by federal and Bavarian state authorities. NHR@FAU hardware is partially funded by the German Research Foundation (DFG) -- 440719683.  
The Quantum Monte Carlo code used to carry on the calculations is available on GitHub \cite{superlat_github}. The figures in this study use Fabio Crameri's scientific colormaps \cite{Crameri2023,Crameri2020}.

\end{acknowledgments}

\appendix

\section{Convergence tests}
In order to exclude that finite-size effects and any systematics coming from the Trotter decomposition we included data of the plasmonic dispersion relation for a lattice of 24x24 unit cells and the imaginary time data of the density-density correlation functions for four exemplarily points in the Brillouin-zone. \\
The dispersion, fitted to $C k^m$, closely follows the square-root dependence, consistent with the results obtained for the smaller lattice with 18x18 unit cells (c.f. Fig.\ref{fig:plasmons_all_model_dispersion} (a)), and thus demonstrating that the plasmonic features are stable with respect to system size and that further lattice enlargement is not necessary to reproduce qualitatice features of the dispersion relaiton.
 \begin{figure}[h]
  \centering
\includegraphics[width=0.45\textwidth , angle=0]{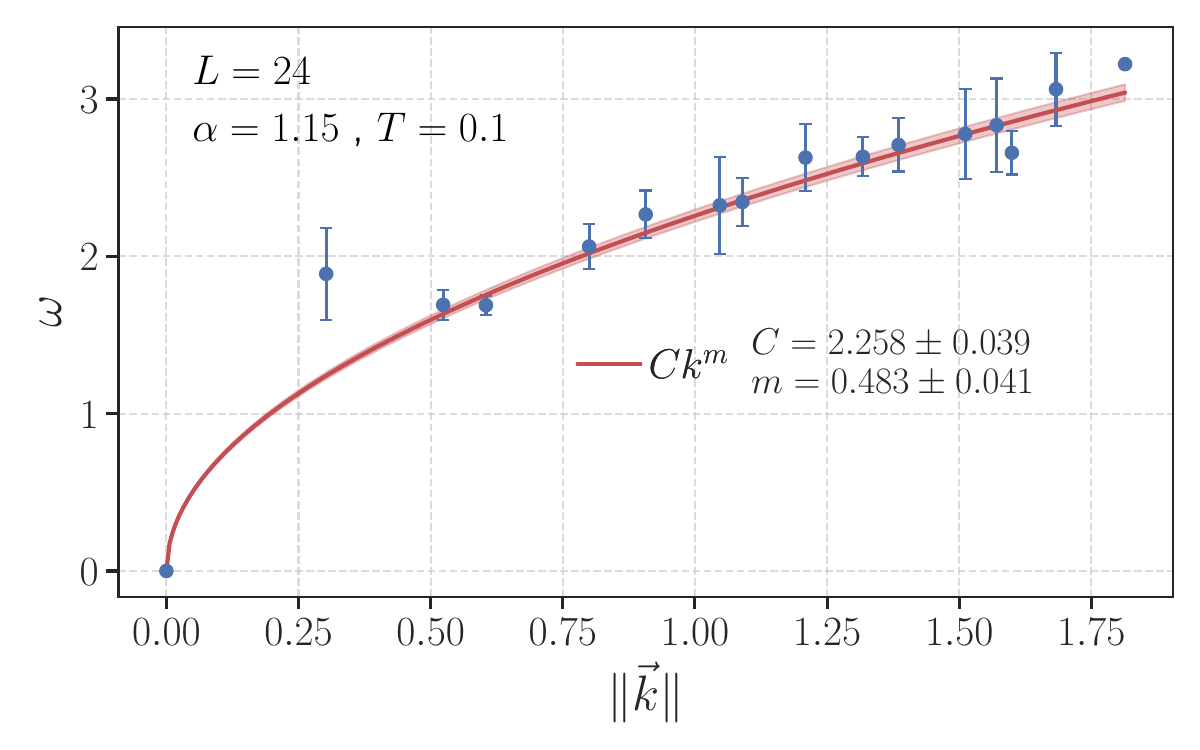}
         \caption{Dispersion relation for a lattice of 24x24 unit cells simulated with the Full Coulomb setup of Eq.~\eqref{eq:Coulomb_law}. }
    \label{fig:Dispersion24x24}
\end{figure}

\begin{figure}[h]
  \centering
\includegraphics[width=0.45\textwidth , angle=0]{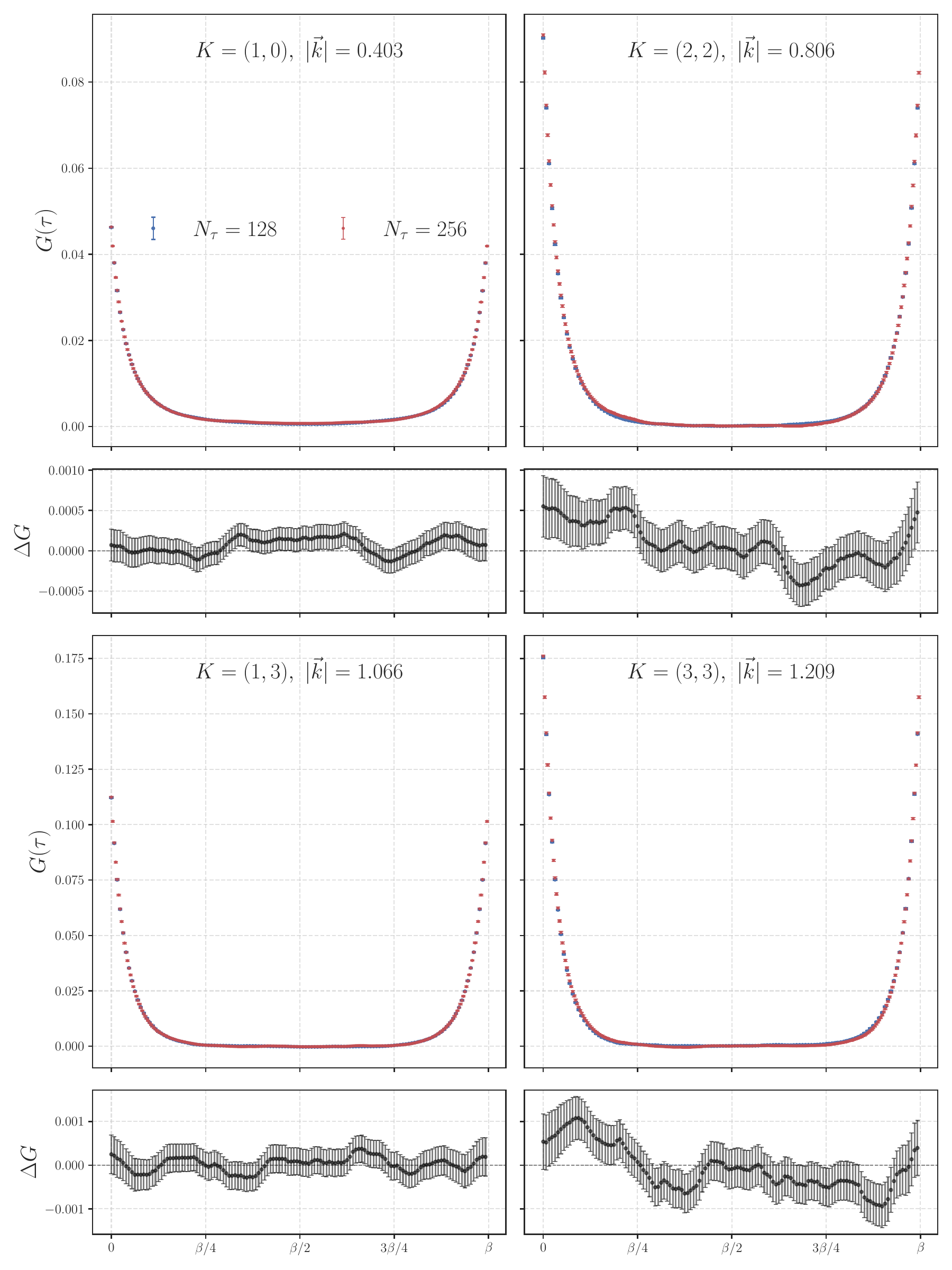}
         \caption{Comparison of the imaginary time data of the density-density correlation function between data generated at $N_t = 128$ and $N_t = 256$. Interaction setup is full Coulomb at $T=0.1$. Below each imaginary time figure we also show the difference between correlators. The data matches exactly within the errorbars. }
    \label{fig:DtauConvergence}
\end{figure}

In addition, the imaginary‑time density–density correlator at various points in the BZ in Fig.~\ref{fig:DtauConvergence} yields consistent results for different numbers of time slices demonstrating that the Euclidean time correlators are also converged with respect to the Trotter discretization.

\bibliography{plasmons_bib}

\end{document}